\documentclass[aps,prx,twocoloumn,reprint,citeautoscript,]{revtex4-1}

\usepackage[colorlinks=true,citecolor=blue]{hyperref}
\hypersetup{colorlinks=true,citecolor=blue,linkcolor=red,urlcolor=blue}

\usepackage{amsmath}
\usepackage{graphicx}
\usepackage{color}
\usepackage{tabularx}
\graphicspath{{.}{./figs/}}
\usepackage[left,pagewise]{lineno}
\begin{document}
\title{Unveiling hidden charge density waves in single-layer NbSe$_2$ by impurities}
\author{Fabrizio Cossu$^{1}$}\email{fabrizio.cossu@apctp.org}
\author{Ali G.\ Moghaddam$^{2}$}\email{agorbanz@iasbs.ac.ir}
\author{Kyoo Kim$^{3}$}
\author{Hassan A.\ Tahini$^{4,5}$}
\author{Igor Di Marco$^{1,6,7}$}
\author{Han-Woong Yeom$^{7,8}$}
\author{Alireza Akbari$^{1,3,7,2}$}\email{alireza@apctp.org}
\affiliation{$^1$Asia Pacific Center for Theoretical Physics, Pohang, Gyeongbuk 790-784, Korea}
\affiliation{$^2$Department of Physics, Institute for Advanced Studies in Basic Sciences (IASBS), Zanjan 45137-66731, Iran}
\affiliation{$^3$Max Planck POSTECH Center for Complex Phase Materials, POSTECH, Pohang 790-784, Korea}
\affiliation{$^4$Department of Applied Mathematics, Research School of Physics and Engineering, Australian National University, Canberra 0200, Australia}
\affiliation{$^5$Integrated Materials Design Centre (IMDC), School of Chemical Engineering, UNSW Australia, Sydney, NSW 2052, Australia}
\affiliation{$^6$Department of Physics and Astronomy, Uppsala University, Box 516, SE-75120, Uppsala, Sweden}
\affiliation{$^7$Department of Physics, POSTECH, Pohang, Gyeongbuk 790-784, Korea}
\affiliation{$^8$Institute for Basic Science Korea, Center for Artificial Low Dimensional Electronic Systems Pohang 790784, South Korea}
%
%
\date{\today}
%
\begin{abstract}
 We employ {\it{ab-initio}} calculations to investigate the charge density waves in single-layer NbSe$_2$, and we explore how they are
 affected by transition metal atoms. Our calculations reproduce the observed orthorhombic phase in single-layer NbSe$_2$ in the clean
 limit, establishing the energy order between three different distorted structures, two consisting of triangular Nb-Nb clusters and a
 third, energetically unfavoured, consisting of hexagonal Nb-Nb clusters. Such energy order, in agreement with known experimental work,
 is reversed by the adsorption of Co and Mn, which favour the formation of hexagonal Nb-Nb clusters; this CDW structure is indeed allowed
 from symmetry point of view but hidden in pure single layers because it is at a higher energy. The other adsorbates, K and Ga, still
 favour one of the triangular Nb-Nb cluster, while suppressing the other. We report how the energy difference between such distorted
 structure varies with these adsorbates. Furthermore, transition metals induce magnetism and favour the reduction of the symmetry of
 the charge density distribution.
\end{abstract}

%
\maketitle
%
\section{Introduction}
\label{Introduction}
  Over the past decade, synthesis and exploration of atomically thin two-dimensional (2D) materials have almost revolutionised our common understandings
 of condensed matter systems and opened a new era in nanosciences \cite{novoselov-PNAS2005,xu_ms-ChemRev2013,choi-mattoday2017}. In particular, 2D
 materials usually show drastically different electronic properties compared to their corresponding bulk structures composed of van der Waals (vdW)
 coupled atomic layers \cite{butler-ACSNano2013,geim-Nature2013,novoselov-Science2016}. Transition metal dichalcogenides (TMDCs) are among vdW layered
 materials which show a wide range of interesting phenomena and applications due to the tunability of their electronic structures
 \cite{splendiani-NanoL2010.4,mak-PRL.105.136805,kuc-PRB.83.245213,wang_qh-nnano2012}. Moreover, they can host exotic phases, such as superconductivity
 (SC), charge density waves (CDW), and even topologically non-trivial states
 \cite{chhowalla-NatChem2013,frindt-PRL.28.299,wilson-AdvPhys2001,gong-SciRep2017,qi-ncomm2016}. Strongly correlated phases in bulk TMDCs have been well
 studied for decades, and recent observations of CDWs and Ising SC \cite{lu_jm-Science2015} in few-layer films have revived interest in these materials
 \cite{xi-nnano2015,xi-nphys2016,ugeda-nphys2016,xi-PRL.117.106801,zhu-ncomm2016}.
\par
  Among TMDCs, NbSe$_2$ has been considered as a prototype material for investigation of CDW orders and SC \cite{bevolo-JAP1974,corcoran-JPCM1994}.
 Bulk 2H-NbSe$_2$ hosts a CDW phase with a $3\times3$ periodicity below 33K which can coexist with an s-wave superconducting phase below
 7K \cite{wilson-PRL.32.882,moncton-PRL.34.734,revolinsky-JPCS1965,rahn-PRB.85.224532,castroneto-PRL.86.4382}, and it is known for its high
 magnetic anisotropy \cite{galvis-ncommPhys2018}. By reducing its thickness down to single layers, the SC weakens but still survives, whereas
 the CDW transition temperature increases \cite{frindt-PRL.28.299,xi-nnano2015,ugeda-nphys2016}. The SC-CDW coexistence in NbSe$_2$ is due to
 a momentum dependent gap opening \cite{rahn-PRB.85.224532,zheng-PRB.97.081101} accompanied by an electronic reconstruction over a wide energy
 range \cite{arguello-PRB.89.235115}, leaving enough electronic states available for the superconducting transition. Such SC-CDW coexistence
 is maintained with charge doping, and their order parameters vary in the same fashion \cite{xi-PRL.117.106801}. The origin of the CDWs has
 been intensively debated for some TMDCs, including NbSe$_2$, mostly because of controversies over the role of Fermi-surface nesting
 \cite{rice-PRL.35.120,liu_r-PRL.80.5762,straub-PRL.82.4504,inosov-NJP2008,rossnagel-PRB.64.235119,rossnagel-PRB.76.073102,johannes-PRB.73.205102,johannes-PRB.77.165135}.
 Nevertheless, in recent years theoretical and experimental evidences has been accumulated in support of momentum-dependent electron-phonon
 coupling as a key mechanism in the formation of CDWs \cite{calandra-PRB.80.241108,valla-PRL.92.086401,zhu-PNAS2015,arguello-PRL.114.037001}.
 Turning to 2D single-layers, the lack of inversion symmetry, the disappearance of the coterminous of vdW interactions and the interplay with
 many-body strong-correlation effects may lead to the aforementioned drastic changes \cite{kotov-RMP.84.1067,guinea-ANDP2014,mak-nmat2012}. 

  On the other hand, the structure of CDWs in bulk and single-layer NbSe$_2$ is controversial \cite{skripov-SSC1985.53,malliakas-JACS2013,silvaguillen-2DMat2016}. While
 a recent detailed experimental work reported evidence of a $3\times3$ commensurate modulation of the crystal structure, the case of the single-layer still needs to be clarified
 \cite{calandra-PRB.80.241108,malliakas-JACS2013}. In addition, impurities or gate doping can play a major role on the CDW behaviour. For example, long-range CDW phase
 coherence can be suppressed in NbSe$_2$ by a moderate percentage of Co or Mn intercalated at the surface \cite{chatterjee-ncomm2015}, Na intercalated in a bi-layer
 \cite{lian-PRB.96.235426} and electron doping \cite{xi-PRL.117.106801}, whereas it can be increased by hole doping \cite{xi-PRL.117.106801}. Besides, Bi adsorption can
 also lead to a transition to a stripe phase \cite{fang-ScienceAdv2018}, previously observed in accidentally doped samples \cite{soumyanarayanan-PNAS2013}.

  In this work, performing an exhaustive first principle calculations, we reveal possible structures of CDWs, particularly at the presence of certain types of
 impurities. Among three different modulated structures, those two with triangular Nb-Nb clusters are found to be energetically favoured in clean NbSe$_2$ 2D
 sheet. These results are consistent with experimental evidence \cite{skripov-SSC1985.53,malliakas-JACS2013}. As a key finding we demonstrate that CDWs with
 hexagonal modulation can be established by adsorption of certain atoms such as Co and Mn. This type of CDW phase is in fact hidden in pristine NbSe$_2$ because
 it has a higher energy compared to triangularly-modulated CDWs. In addition, it is uncovered that the presence of the transition metal impurities induces
 magnetism and promotes modulated phases with reduced symmetry of the charge density distribution compared to pure CDW structures. Other types of metallic
 ad-atoms, namely K and Ga, allow the same ground state as pristine single-layer NbSe$_2$, but Ga also supports the hexagonally modulated structure. The
 current manuscript aims at steering future research towards a new interpretation of the experimental evidence on the effect of impurities
 \cite{fang-ScienceAdv2018,chatterjee-ncomm2015,soumyanarayanan-PNAS2013}, and represents an important contribution in the field of the interplay between CDWs
 and SC, as a recent work strongly points out \cite{cho-ncomm2018}.

 This paper is organised as follows. In Sec.\ \ref{comp}, we will briefly introduce the computational method based on density functional theory
 which is used for investigation of CDW phases. Thereafter, we go through the results of \emph{ab-initio} calculations for CDWs in pristine
 NbSe$_2$, Sec.\ \ref{pristine}, where the relaxed CDW structures and the profiles of charge densities are presented both in real space and in
 Fourier transformed form. The core of our work is found in Sec.\ \ref{impure}, which shows the CDWs in the presence of various impurities. In
 particular, we show how the energy hierarchy of the CDWs can be different from the pristine system and a hidden order arise as a new ground
 state by adding Co or Mn ad-atoms. Finally, after a discussion over the results, the conclusions are presented in Sec.\ \ref{conc}.

\section{Computational details}
\label{comp}
  Our results are obtained by means of density-functional theory (DFT). We employ the projected augmented wave (PAW) method with Perdew-Burke-Ernzerhof (PBE) pseudopotentials,
 as implemented in the Vienna {\itshape{ab initio}} Simulation Package (VASP) \cite{blochl-PRB.50.17953,kresse-PRB.59.1758}. Accordingly, the exchange-correlation functional
 is treated in the generalised gradient approximation in the PBE parametrisation \cite{perdew-PRL.77.3865,perdew-PRLerratum.78.1396}.
 The basis set consists of plane waves, with the explicit treatment of 13, 6, 9, 15, 9 and 13 valence electrons for Nb, Se, Co, Mn, K and Ga states, respectively.
 As previously suggested \cite{wehling-PRB.84.235110}, standard local and semi-local exchange-correlation functionals may not offer a proper description of the partially
 filled $3d$ and $4f$ shells of TM adatoms on 2D materials. Therefore, calculations involving Co and Mn are performed in the DFT+U approach,
 using the rotationally invariant formulation of Lichtenstein {\itshape{et al.}} \cite{liechtenstein-PRB.52.R5467}.
 As in ref.\ \onlinecite{wehling-PRB.84.235110}, we use a generalised value of $U = 4.00$ eV and
 $J = 0.90$ eV for the $3d$ orbitals of both Co and Mn. Small variations of these parameters (within a reasonable
 range) are unlikely to change the physical picture outlined in the present paper. For sake of completeness we 
 analyse this issue for Co adatoms, in the Appendix.
 For all calculations, the cutoff energy of the plane waves is 400 eV, while the energy tolerance on the electronic loops for the
 relaxation and for the electronic properties are set to 10$^{-6}$ eV and 10$^{-7}$ eV, respectively; a conjugate gradient algorithm
 is employed for structural relaxation. Structures are considered relaxed when the forces on each atom are smaller than 2 meV/\AA.
 The simulations are run in $3\times3\times1$, $6\times6\times1$ and $9\times9\times1$ replicas of the NbSe$_2$ single-layer unit cell. After convergence tests
 on the {\bf k}-meshes for the $3\times3\times1$ and $6\times6\times1$ replicas were performed, $15\times15\times1$, $7\times7\times1$ and $5\times5\times1$
 grids of {\bf k}-points were used to sample their Brillouin Zones for the total energy calculations; $11\times11\times1$ and $5\times5\times1$ {\bf k}-meshes
 were used for the partial charge density calculations in the $6\times6\times1$ and $9\times9\times1$ replicas, respectively; a $45\times45\times1$
 {\bf k}-mesh was used for the DOS calculations in the $3\times3\times1$ replica.
 In modelling metal adsorption on NbSe$_2$, the concentration of one ad-atom in a $6\times6\times1$ replica of the unit cell was adopted, corresponding to 0.0278
 impurities/f.u., which allows for a description of a $3\times3\times1$ CDW while minimising the interaction between impurities and their images. The adsorbates
 taken into account in the present study are Co, Mn, K and Ga.

\begin{figure*}[t]
\centering
 {\includegraphics[trim = 0 0 0 0,width=\textwidth,clip]{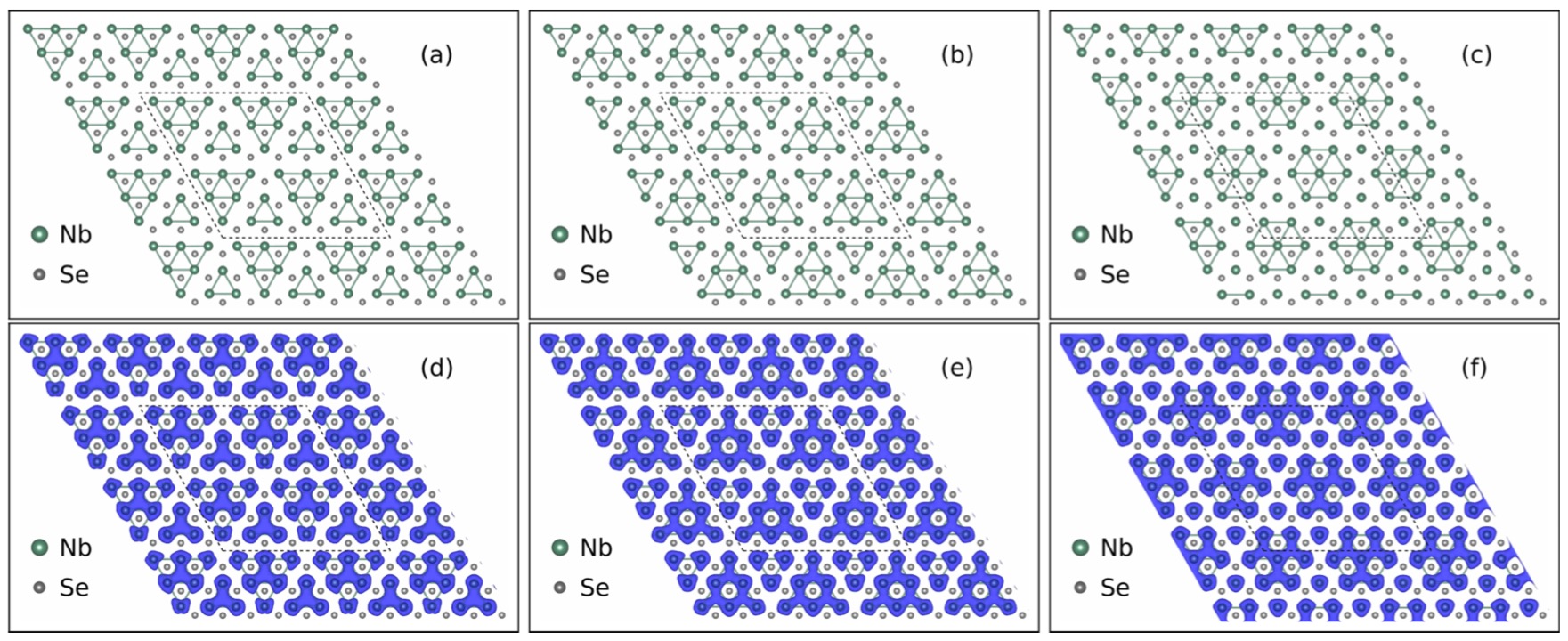}}
\caption{(Colour online) First row: relaxed CDW structures for pristine NbSe$_2$ (top view):
 T-U (a), T-C (b) and HX (c) CDWs (see text for details); Second row: charge densities of
 the respective CDW structures. Volumetric data are represented as blue surfaces enclosing
 points whose electronic density is greater than or equal to 0.0075 electrons/Bohr$^3$.
 Atoms are represented by spheres, as illustrated in the legends (magnified); Nb-Nb bonds
 shorter than the equilibrium distance (3.45 \AA) are represented by solid lines, in order
 to help visualising the CDW structure pattern. Dashed lines mark the supercells borders.}
\label{prstStrcts-fig}
\end{figure*}
\begin{figure*}[t]
\centering
 {\includegraphics[trim = 0 0 0 0,width=\textwidth,clip]{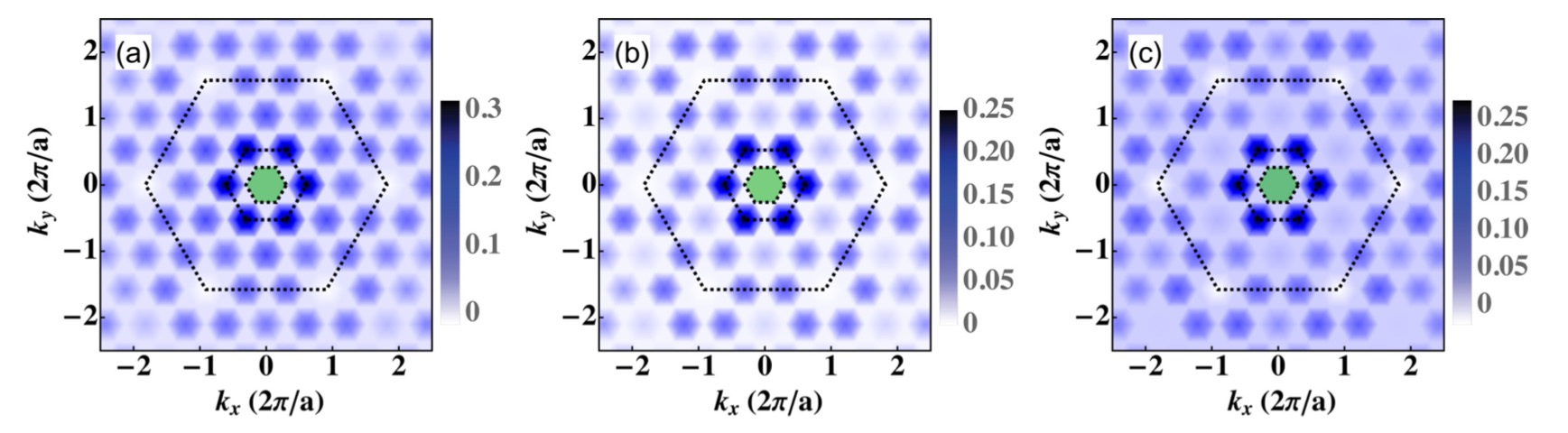}}
\caption{(Colour online) Differences between the Fourier Transform (FT) of the charge density
 distributions for the pristine CDW structures and the non-modulated structure. The plots of
 T-U (a), T-C (b) and HX (c) CDWs charge density correspond to those in Fig.\ \ref{prstStrcts-fig}.
 The large dashed hexagon with vertices at ${\left|\mathbf q\right|} = \frac{2\pi}{a}$
 (with $a = 3.45$ \AA) marks the characteristic Bragg peaks; the medium size dashed hexagon with
 vertices at ${\left|\mathbf q\right|} = \frac{2\pi}{3a}$ marks the (shorter) CDW peaks; the small
 green shaded hexagon (${\left|\mathbf q\right|} = \frac{2\pi}{6a}$) maps points beyond the supercell
 borders.}
\label{FFTprst-fig}
\end{figure*}

\section{CDW phases in pristine single layers}\label{pristine}
  We begin with the study of the single-layer NbSe$_2$ in a $3\times3\times1$ supercell, which is the minimal size cell for a CDW in NbSe$_2$. Pristine single-layer NbSe$_2$
 is known to be metallic, non magnetic and hosts CDWs below 145 K \cite{wilson-AdvPhys1969,moncton-PRL.34.734,wilson-AdvPhys2001}. Our models include three structures
 obtained according to existing work \cite{skripov-SSC1985.53,malliakas-JACS2013,xi-nnano2015}, which are shown in Fig.\ \ref{prstStrcts-fig}; the Nb atoms cluster
 in triangular patterns - see Figs.\ \ref{prstStrcts-fig}(a~and~b) - and hexagonal ones - see Fig.\ \ref{prstStrcts-fig} (c); the two triangular patterns differ by the
 position of the Se atoms with respect to the triangle composed by the Nb-Nb bonds; accordingly, the CDW structures are named T-U, T-C and HX, respectively (T-U and T-C
 stand for triangle-uncentred and triangle-centred, whereas HX stands for hexagonal). The two structures T-U and T-C are related to each other by a mirror reflection of
 the Nb sublattice. It is also known that the Se-Se bond patterns (analogous to Nb-Nb bond patterns) accompany those of Nb-Nb \cite{silvaguillen-2DMat2016}. The T-U and
 T-C CDWs differ for the Se-Se pattern, see the Appendix for details. Their total energies were compared in $3\times3\times1$ and $6\times6\times1$ replicas of the NbSe$_2$
 unit cell, and suggest that the formation of all of them is favoured. The T-U is the lowest energy CDW structure; calculations in the $3\times3\times1$ supercell yield
 differences of 3.9 meV, 0.5 meV and 1.3 meV/f.u. with respect to the undistorted structure, T-C and HX, respectively. In the $6\times6\times1$ supercell, slight changes
 are observed in the energy differences between the CDW structures: the T-U is favoured by 0.4 meV and 1.1 meV over the T-C and HX, respectively; these differences are
 maintained for a $9\times9\times1$ supercell. The three CDW structures were recently investigated elsewhere on $3\times3\times1$ supercells \cite{lian-NanoL2018.5}.
 Here, we demonstrate our agreement with ref.\ \onlinecite{lian-NanoL2018.5}, and we make use of the results for the pristine to compare the metal-adsorbed NbSe$_2$.

  The effect of the CDW distortions on the density of states (DOS) is analysed in the Appendix. For what concerns the electronic reconstruction following the
 CDW formation, our calculations are in agreement with the literature \cite{shen-PRL.99.216404,calandra-PRB.80.241108}. The charge distributions are computed
 integrating the charge density over the occupied Nb band (the band crossing Fermi level, see the Appendix) and they are shown in Fig.\ \ref{prstStrcts-fig},
 second row. The integration over the whole occupied band simulates a topography retaining the symmetry of the Nb band only. The charge density clusters in
 patches, with different shape and patterns for each CDW; the lowest energy CDWs, T-U and T-C, have three-fold symmetric patterns and the HX CDW has six-fold
 symmetric ones. The patches of T-U are on the vertices of a hexagon (with no element on the centre), whereas those of T-C and HX are placed on the vertices
 and centre of a larger hexagon.

  The Fourier Transform (FT) of the charge density distribution allows to recognise more clearly the symmetry of the modulation patterns and trace their length
 scale. Due to the three-dimensional periodic boundary conditions, the three-dimensional data is originally computed as a function of the $(h, k, l)$ Miller
 indices; the subset with $l = 0$ is analysed to track modulations of the CDW charge distributions only along the plane. Thus, the FT is mapped as a function
 of $h$ and $k$ ($k_x$ and $k_y$ in the relative plots). The computed FT plots of the charge density distributions of the three CDW structures in pristine
 NbSe$_2$ are reported in Fig.\ \ref{prstFFT-fig}, in Appendix A. 
 
 More significant information can be obtained by considering the difference between those plots and the FT plot of the charge
 density distribution of the non-modulated structure, shown in Fig.\ \ref{FFTprst-fig}. In these (and following) plots, the
 vertices of the large hexagon, at ${\left|\mathbf q\right|} = \frac{2\pi}{a}$, mark the position of the characteristic Bragg
 peaks, which are evident in Fig.\ \ref{prstFFT-fig}. The vertices  of the medium size hexagon,
 at ${\left|\mathbf q\right|} = \frac{2\pi}{3a}$, mark instead the position of the shorter peaks associated to the CDW. Both
 hexagons are emphasised with a dashed line, as a guide for the eye. The small green shaded hexagon, whose vertices are at
 ${\left|\mathbf q\right|} = \frac{2\pi}{6a}$, map points beyond the supercell borders, and thus are not meaningful. Due
 to the size of the supercell, the width of the spots denoting Bragg peaks or CDW peaks is large because different but close
 modulation frequencies cannot be resolved, even with a relatively dense k-mesh. Fig.\ \ref{FFTprst-fig} show the relative
 differences between the three CDWs and the non-modulated structure. The CDW main peaks are not sensibly different from plot
 to plot, neither are satellite peaks at $(2/3) \Gamma K$ and at $(1/3) K K'$. However, a sizeable satellite peak appears at
 $2/3 \Gamma M$ in the T-U CDW, but not (as large) in the T-C nor in the HX CDWs. Such configuration of peaks mirrors the
 configuration of the electronic patches in Fig.\ \ref{prstStrcts-fig}, which in the T-U (T-C and HX) are placed at the vertices
 of a small (large) hexagon, without (with) a central element. In fact, two different vectors map equivalent patterns in T-U.
 Naming the lattice vectors of the unit cell $\mathbf a$ and $\mathbf b$, the mapping vectors are $3\mathbf{a}$ (and equivalently
 $3\mathbf{b}$) and $\mathbf a + \mathbf b$; the latter does not map equivalent patches in
 T-C nor in HX.

\begin{table}[t]
\centering
 \caption{Energetics and magnetism for M$\vert$NbSe$_2$ (M = Co, K, Ga, Mn).
 Energy differences are computed with respect to the lowest configuration
 and are expressed in meV. Magnetic moments are expressed in $\mu_B$.}
 \label{tab-ads}
 \begin{tabular}{r|r||r|r|r}
                     &        & $\Delta$E & $\mu^{TM}$ & $\mu_{tot}$ \\
 \hline\hline
                     & hollow &    16     &     1.9    &      2.2    \\
                  Co & top Nb &     0     &     1.9    &      2.0    \\
                     & top Se &  3551     &     2.0    &      3.4    \\
  \hline
                     & hollow &    91     &     4.4    &      5.8    \\
                  Mn & top Nb &     0     &     4.4    &      4.3    \\
                     & top Se &  1291     &     4.9    &      4.1    \\
  \hline
                     & hollow &     4     &     0.0    &      5.9    \\
                   K & top Nb &     0     &     0.0    &      0.0    \\
                     & top Se &   241     &     0.0    &      1.2    \\
  \hline
                     & hollow &   183     &     0.0    &      5.9    \\
                  Ga & top Nb &     0     &     0.0    &      0.0    \\
                     & top Se &   605     &     0.0    &      0.2    
 \end{tabular}
\end{table}
\begin{table}[t]
\centering
 \caption{Energetics for the CDWs in M$\vert$NbSe$_2$ (M = Co, Mn, K, Ga); the energy for each column
 (adsorbate) is given as differences with respect to the ground state. Different position and resulting
 structure are possible for each metal M adsorbed on a CDW; we report those which evolve to structures
 relatively close in energy to the ground state, mentioning the type of structure, when it is similar to
 one of the three pristine CDWs, as well as the energy difference.
 The difference in energy in the 9x9 supercell (1 Co) is 0.3 meV/f.u. in favour of the HX CDW.}
 \label{tab-cdwia} 
 \begin{tabular}{lr||lr|lr|lr|lr}
   \multicolumn{2}{c}{pristine}  & \multicolumn{2}{c|}{Co} & \multicolumn{2}{c|}{Mn} &  \multicolumn{2}{c}{K}  & \multicolumn{2}{|c}{Ga}   \\
 \hline\hline                                                                 
                    T-U & (0.0)  & T-U        &      (0.7) & T-U        &      (1.5) &  T-Uh      &     (0.0)  & T-U        &      (0.0)   \\
                        &        & T-U        &      (2.6) &            &            &  T-UN      &     (0.2)  &            &              \\
\hline                                                                        
                    T-C & (0.4)  & HX-A       &      (0.0) & HX-A       &      (0.0) &  T-CN      &     (1.5)  & HX-A       &      (0.1)   \\
                        &        &            &            &            &            &  T-Ch      &     (1.3)  &            &              \\
\hline                                                                        
                     HX & (1.1)  & HX-S       &      (0.0) & HX-S       &      (0.0) &  HX-S      &     (0.7)  & HX-S       &      (0.0)   \\
                        &        & HX-A       &      (0.0) & HX-A       &      (0.0) &            &            &            &              
 \end{tabular}
\end{table}
\begin{table}[t]
\centering
 \caption{Energetics and magnetism for M$\vert$NbSe$_2$ (M = Co, K, Ga, Mn).
 Energies are given as differences with respect to the lowest energy solution
 and are expressed in meV; magnetic moments are expressed in $\mu_B$.}
 \label{tab-adscdw} 
 \begin{tabular}{r|r||r|r|r}
                     &        & $\mu^{TM}$ & $\mu_{tot}$ &  $E^{ad}$  \\
 \hline\hline
                     &   T-U  &     1.9    &      2.0    &   3.440    \\
                  Co &  HX-A  &     1.9    &      2.0    &   3.504    \\
                     &  HX-S  &     1.9    &      1.9    &   3.504    \\
 \hline
                     &   T-U  &     4.4    &      4.4    &   2.472    \\
                  Mn &  HX-A  &     4.4    &      4.4    &   2.559    \\
                     &  HX-S  &     4.4    &      4.3    &   2.561    \\
 \hline
                     &   T-Uh &     0.0    &      0.0    &   3.045    \\
                   K &   T-UN &     0.0    &      0.0    &   3.038    \\
                     &   HX-S &     0.0    &      0.1    &   3.060    \\
 \hline
                     &   T-U  &     0.0    &      0.0    &   2.907    \\
                  Ga &  HX-A  &     0.0    &      0.0    &   2.946    \\
                     &  HX-S  &     0.0    &      0.0    &   2.944    \\
 \end{tabular}
\end{table}

\section{CDW phases in the presence of impurities}\label{impure}

\begin{figure*}[t]
\centering
 {\includegraphics[trim = 0 0 0 0,width=\textwidth,clip]{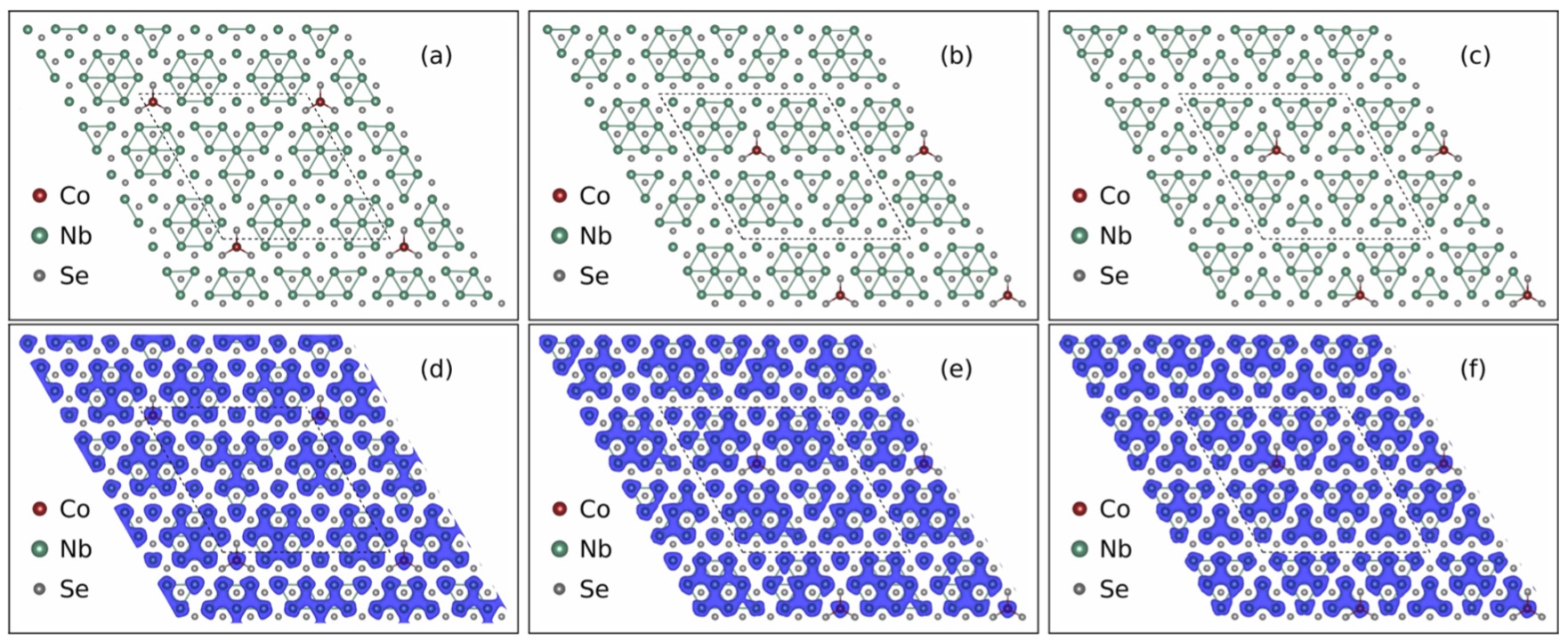}}
\caption{(Colour online) First row: ground state Co$\vert$NbSe$_2$ HX-S (a), HX-A (b) and T-U (c)
 CDW structure; Second row: charge densities of the respective CDW structures. The isosurface
 value for the volumetric data is set in agreement with Fig.\ \ref{prstStrcts-fig}. Atoms are
 represented by spheres, as illustrated in the legends; Nb-Nb bonds shorter than the equilibrium
 distance (3.45 \AA) are represented by solid lines, to help visualising the CDW structure pattern.
 Dashed lines mark the supercells borders.}
\label{CoStrcts-fig}
\end{figure*}
\begin{figure*}[t]
\centering
 {\includegraphics[trim = 0 0 0 0,width=\textwidth,clip]{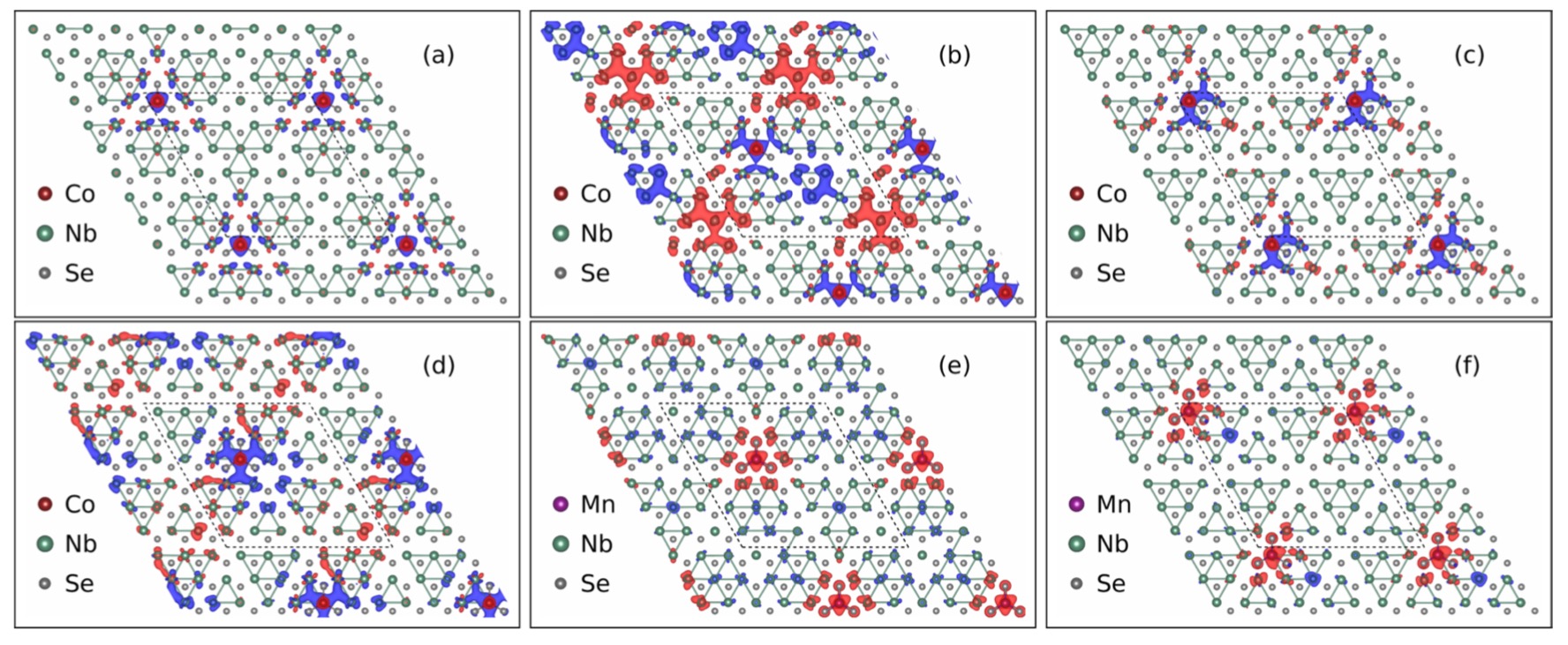}}
\caption{(Colour online) Magnetisation densities of Co$\vert$NbSe$_2$ HX-S (a), Co$\vert$NbSe$_2$
 HX-A (b), Co$\vert$NbSe$_2$ T-U at 1 (c), Co$\vert$NbSe$_2$ T-U at 3 (d), Mn$\vert$NbSe$_2$ HX-A (e)
 and Mn$\vert$NbSe$_2$ T-U at 1 (f). The magnetisation densities in (c) and (d) are different although
 they have virtually the same structure and total energy; the magnetisation in (c) is juxtaposed to (f)
 to compare structures with the same adsorption site. Volumetric data are represented as blue (red)
 surfaces enclosing points whose magnetisation density is greater than or equal to 0.0075 (smaller
 than or equal to -0.0075) $\mu_B$/Bohr$^3$. Atoms are represented by spheres, as illustrated in
 the legends; Nb-Nb bonds shorter than the equilibrium distance (3.45 \AA) are represented by solid
 lines to help visualising the CDW structure pattern. Dashed lines mark the supercells borders.}
\label{CoMnMgn-fig}
\end{figure*}
\begin{figure*}[t]
\centering
 {\includegraphics[trim = 0 0 0 0,width=\textwidth,clip]{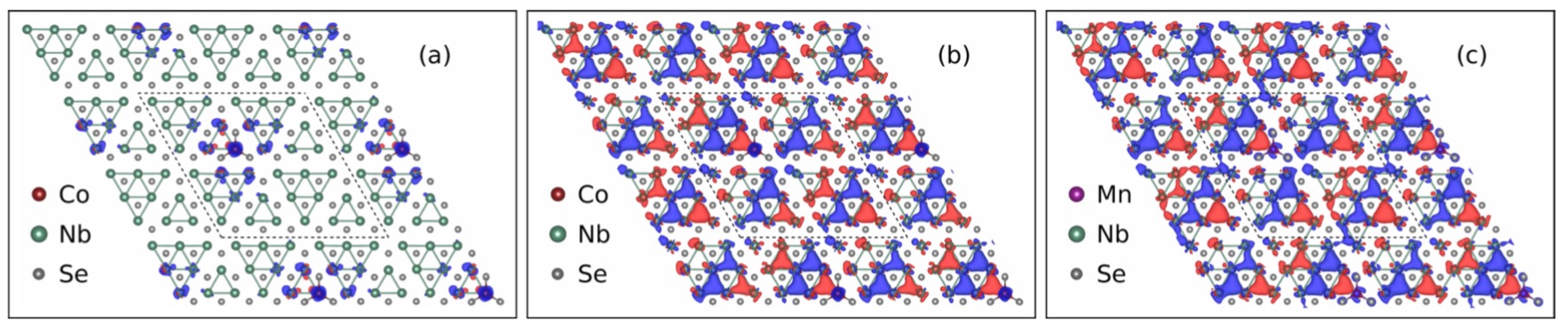}}
\caption{(Colour online) Differences between charge density distributions. Co$\vert$NbSe$_2$
 T-U minus NbSe$_2$ T-U (a), Co$\vert$NbSe$_2$ HX-A minus NbSe$_2$ T-U (b) and Mn$\vert$NbSe$_2$ HX-A
 minus NbSe$_2$ HX (c). Volumetric data are represented as blue (red) surfaces enclosing points whose
 electronic density is greater than or equal to 0.0025 (smaller than or equal to -0.0025)
 electrons/Bohr$^3$. Atoms are represented by spheres, as illustrated in the legends; Nb-Nb bonds
 shorter than the equilibrium distance (3.45 \AA) are represented by solid lines to help visualising
 the CDW structure pattern. Dashed lines mark the supercells borders.}
\label{CoMnChgDiffs-fig}
\end{figure*}
\begin{figure*}[t]
\centering
 {\includegraphics[trim = 0 0 0 0,width=\textwidth,clip]{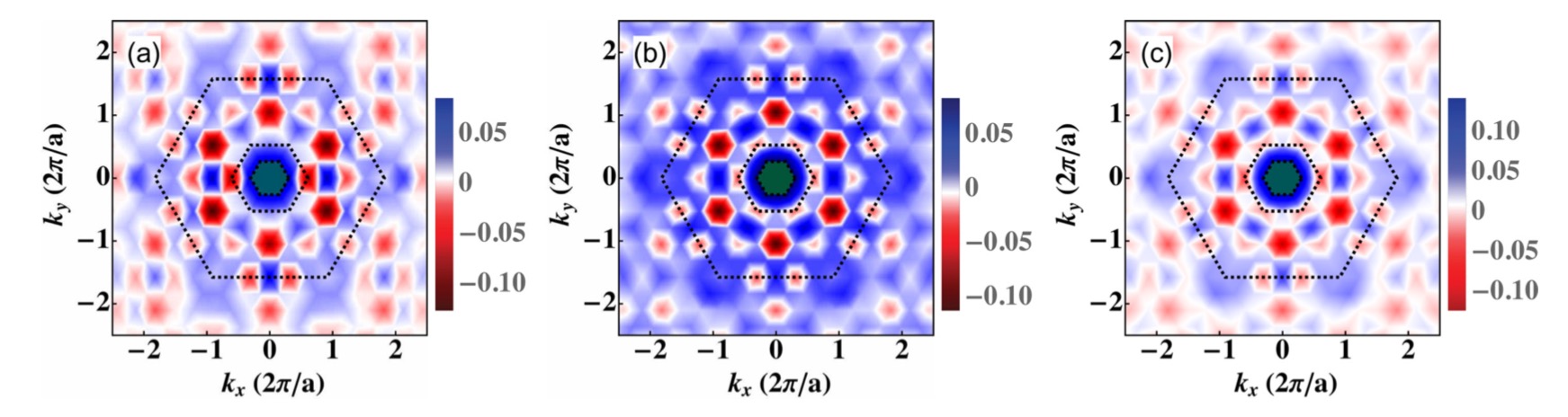}}
\caption{(Colour online) Differences between FT of the charge density
 distribution of Co-adsorbed and Mn-adsorbed NbSe$_2$ CDWs and the T-U CDW in
 pristine NbSe$_2$. Ground state Co$\vert$NbSe$_2$ HX-A (a), Co$\vert$NbSe$_2$
 HX-S (b) and Mn$\vert$NbSe$_2$ HX-A (c). The dashed hexagons mark the same
 regions as in Fig.\ \ref{FFTprst-fig} (see caption).}
\label{FFTCoMn-fig}
\end{figure*}

  Adsorption of the adsorbates Co, Mn, K and Ga on a fully symmetric structure induces structural distortions breaking the symmetry of the NbSe$_2$ layer
 according to the site: the adsorption on the Nb site, hollow site and Se site induces CDWs of HX type (i.e. having similar Nb-Nb distance patterns to the
 pristine HX), T-U type and T-C type, respectively. By total energy calculations, the likelihood of the adsorption sites is analysed. The energies
 for different adsorption sites of the investigated ad-atoms are reported in Table \ref{tab-ads}.  These results are consistent with
 those reported in a recent work on MoS$_2$ \cite{wei-PRB.95.075419}.
 The preferred adsorption site is on top of a Nb atom for Co, K, Ga and Mn. The competition between hollow site
 and Nb site is strong in the case of K (4 meV difference). The Se site remains the most unfavoured for single atoms. However, it competes with Nb for larger
 molecules. In order to highlight this trend, we computed the energy difference for Co-(OH)$_2$, which can be considered as a prototype of a small molecule
 and a possible impurity. In the case of a single Co, the Nb site is preferred to the hollow site and Se site by 16 meV and 3551 meV, respectively; in the case
 of Co-(OH)$_2$, the corresponding energy difference are $\sim$~670 meV and 166 meV, suggesting a hindrance to the Co-Se bonds and Co-Nb bonds or a decrease
 of the Co charge state and in turn of its coordination. In summary, single Co is preferably adsorbed on the Nb site, with the hollow site relatively close
 in energy; large molecules favour adsorption on the Se site and unfavour adsorption at the hollow site, a trend in line with combined theoretical and
 experimental results \cite{kezilebieke-NanoL2018.4}. After establishing the preferred adsorption sites on the NbSe$_2$ non modulated structure for each
 adsorbate by total energy calculations, the adsorption on the different CDW structures is modelled (it involves different inequivalent adsorption sites
 due to the lower symmetry).

\begin{table}[th]
\centering
 \caption{Occupancies on Co and Mn $3d$ orbitals, which split according to a trigonal prismatic
 environment, namely $\mathit{e''}$, $\mathit{e'}$ and $\mathit{a'_{1}}$, where $\mathit{a'_{1}}
 = d_{z^2 - 3r^2}$. The coefficients $u_{\uparrow}$, $v_{\uparrow}$, $u_{\downarrow}$ and
 $v_{\downarrow}$ are .90, .43, .62 and .78 for Co and .99, .16, .78 and .62 for Mn, respectively.}
 \label{tab-occTM}
 \begin{tabular}{r||c|c|c|c}
                           &           \multicolumn{2}{c|}{Co}           &            \multicolumn{2}{c}{Mn}            \\
 \hline
                           & $\sigma = \uparrow$ & $\sigma = \downarrow$ & $\sigma = \uparrow$ & $\sigma = \downarrow$  \\
 \hline\hline
  $\mathit{e'_{1,\sigma}}$ &        0.93         &          0.07         &        0.93         &          0.03          \\
         $\mathit{a'_{1}}$ &        0.90         &          0.79         &        0.92         &          0.08          \\
 $\mathit{e''_{1,\sigma}}$ &        0.95         &          0.92         &        0.93         &          0.09
 \end{tabular}
\end{table}

\subsection{Adsorption of Co}

  Upon adsorption of Co, the energy and state of CDWs are modified. In general, the solutions are a combination of the states found in the pristine system.
 Pure solutions of the T-U CDW are found and are referred to as such in the remainder. The HX CDW solutions are found in mixed states or pure states; those
 with lowest energy are grouped according to the symmetry of their structures and charge densities, and named HX-S and HX-A, for symmetric and asymmetric
 ones, respectively. The T-C CDW solutions are found only mixed (with the HX-A CDW). The HX-S, HX-A and T-U CDWs structures are shown in Figs.\
 \ref{CoStrcts-fig} (a), (b) and (c), respectively. The HX-S features the characteristic hexagonal patches of the pristine HX CDW together with a tri-fold
 symmetric star of Nb-Nb bonds, compare Figs.\ \ref{CoStrcts-fig} (a) and \ref{prstStrcts-fig} (c). Their charge density distributions are not remarkably
 different from their pristine counterparts, compare Figs.\ \ref{CoStrcts-fig} (d) and \ref{prstStrcts-fig} (f). The flatness of the (multi-dimensional)
 potential energy surface allows adsorption on different sites of the underlying CDW structure to give different solutions. As Table \ref{tab-cdwia} reports,
 the HX-A CDW results from the relaxation of the adsorption of Co on the vertex of the large triangle of a T-C CDW structure, compare Fig.\ \ref{prstStrcts-fig} (b),
 and a mixing between the T-C and the HX CDWs occurs. The patches in the charge density distribution, Fig.\ \ref{CoStrcts-fig} (e), recall those in both the
 pristine T-U and the pristine T-C, Figs.\ \ref{prstStrcts-fig} (e) and (f), supporting the previous observation. Finally, the structure and charge density
 distribution of the T-U solution are virtually identical to those in the pristine T-U CDW, with minor differences around the adsorption site, compare Figs.\
 \ref{CoStrcts-fig} (f) and \ref{prstStrcts-fig} (a).

  The magnetic density distributions exhibit antiferromagnetic coupling between Co and Nb in the occupied Nb band; moreover the magnetisation around Co in that
 energy range is opposite to the total magnetisation on Co, compare Fig.\ \ref{CoMnMgn-fig}, first row, with Table \ref{tab-adscdw}. The magnetic moment on Co
 (value integrated over all the occupied states) is around 2.0 $\mu_B$ for every CDW, and the total magnetisation over the whole NbSe$_2$ layer vanishes. The
 same observation is valid for Mn, see Table \ref{tab-adscdw}, and the discussion in the Appendix. The modulation of the magnetisation density
 in the HX-A CDW is larger than that in the HX-S (and that in T-U) CDW, compare Figs.\ \ref{CoMnMgn-fig} (a) and (b), first row, suggesting that larger mixing
 of different CDWs supports the formation of a spin density wave (SDW).
  Moreover, a different modulation of the magnetisation densities of Co$\vert$NbSe$_2$ T-U with Co at different adsorption CDW site is observed, compare
 Figs.\ \ref{CoMnMgn-fig} (c) and (d). These two structures are at the same energy, suggesting that a magnetic order transition is still incipient. The
 magnetisation data obtained by a site-by-site analysis point to a ferromagnetic coupling, in agreement with ref.\ \onlinecite{zhou-ACSNano2012}, where an
 incipient magnetic transition is achieved by tensile strain.

 The variation of the charge modulation can be analysed looking at the difference in the charge density distribution between
 Co$\vert$NbSe$_2$ CDWs and the pristine CDWs in the direct space, see Fig.\ \ref{CoStrcts-fig}. In these plots, blue (red)
 lobes denote injection (depletion) of charge with respect to the pristine CDW charge distribution. The adsorption of Co modifies
 the T-U CDW only in the neighbourhood of the adsorption site, see Fig.\ \ref{CoMnChgDiffs-fig} (a), as expected from Fig.\
 \ref{CoStrcts-fig} (f); charge depletion (injection) occurs out-of-plane (in-plane) in correspondence to the Nb at the adsorption
 site and in a more complex pattern on the surrounding Nb atoms; in general, red lobes point towards the adsorbate. The charge
 difference between the HX-A and the T-U CDWs features a constant charge difference - due to the misalignment between the two
 charge densities (in NbSe$_2$ and Co$\vert$NbSe$_2$) - and an enhancement of in-plane modulations identified by the isolated
 blue lobes in Fig.\ \ref{CoMnChgDiffs-fig} (b); these may be more relevant in comparison to the out-of-plane modulations
 represented by the out-of-plane red lobes on isolated Nb atoms in Fig.\ \ref{CoMnChgDiffs-fig} (c), which refer to Mn-adsorbed
 NbSe$_2$.

 Eliminating charge displacements (due to the different adsorption sites) helps analysing the symmetry of the charge distribution. In analogy to
 what done for the pristine case (Fig.\ \ref{FFTprst-fig}), we do not show the FT plots of the charge density distributions directly. Instead, we
 focus on the differences between these plots and the FT plot obtained for the T-U CDW in the pristine case. In order to identify variations in
 the CDW signal, the marks of Bragg peaks, CDW peaks and the region corresponding to points in the direct space beyond the supercell size are used
 in agreement with Figs.\ \ref{FFTprst-fig}.

  The FT plots relative to Co$\vert$NbSe$_2$ HX-A, Co$\vert$NbSe$_2$ HX-S and Mn$\vert$NbSe$_2$ HX-A are shown in Figs.\ \ref{FFTCoMn-fig} (a)-(c); the FTs shown
 are computed difference with respect to the pristine T-U CDW. In these cases with adsorbates, a peak in the neighbourhood of ${\left|\mathbf q\right|} = 0$ appears,
 because of the background charge (uniformly) injected into the system. All plots in Fig.\ \ref{FFTCoMn-fig} point to a considerable suppression (enhancement) of the
 CDW intensity along $\Gamma K$ at $\left|\mathbf q \right| = 2\pi/3a$ (at $\left|\mathbf q \right| = \pi/a$); such suppression/enhancement is anisotropic for Co HX-A,
 being large along the line $\mathbf k_y = 0$ and small along the other two $\Gamma K$ lines; the other cases are isotropic (Co HX-S) or virtually isotropic (Mn HX-A).
 Furthermore, large depletion of intensity at $\left|\mathbf q \right| = 4\pi/3a$ along $\Gamma M$ as well as at $\left|\mathbf q \right| = 2\pi/3a, 4\pi/3a$ along
 $K K'$, illustrates again the difference between the pristine HX CDW (T-C as well) and the pristine T-U CDW observed above. Finally, intensity enhancement occurs
 also within the small hexagon marking the $3\times3$ CDW peaks (but with no leading ${\mathbf q}$-vector), suggesting a competition between modulations with different
 wavelengths, as previously discussed \cite{calandra-PRB.80.241108,lian-NanoL2018.5}. Indeed, strain-induced modifications of the $\mathbf q$ vector were recently
 found \cite{gao-PNAS2018}. In general, the asymmetric form of the CDW intensity (with respect to pristine T-U) suggests a connection with recently observed stripe
 phases \cite{soumyanarayanan-PNAS2013,fang-ScienceAdv2018,gao-PNAS2018} of which it could be a precursor. The character of the asymmetry in the adsorbates systems is
 further treated in the Appendix, with reference to Fig.\ \ref{otherFFT-fig}.

\subsection{Adsorption of Mn}

  In the case of Mn adsorption, magnetism plays a major role. A high magnetic moment, which correctly describes Mn, favours the Nb site more than in
 the case of Co adsorption (the energy difference between the Nb site and the hollow site is 5 times larger), compare their energy differences in Table
 \ref{tab-ads}. The ordering of the CDW structures, assessed by total energy calculations, follows the same pattern as in the case of Co adsorption,
 having a family of ground state HX CDWs, including a symmetric one and an asymmetric one, a solution of T-U at a higher energy and the absence of a
 T-C solution. The energy difference between the ground state HX CDW and the T-U solution is 1.5 meV/f.u., see Table \ref{tab-cdwia}, which compared
 to the case of Co, it suggests that Mn drives a slightly stronger transition to a HX CDW. The charge density distributions of the HX-S and HX-A show
 no essential difference, being also very similar to that of Co$\vert$NbSe$_2$ HX-S. Therefore, the relative figures are omitted, and the remaining
 discussion is limited to the magnetisation density distributions and the FT of the charge density distributions.

  The magnetisation densities of the HX-S, HX-A the T-U are shown in Fig.\ \ref{CoMnMgn-fig}, second row. The two HX CDWs are similar also in their
 magnetisation density distribution - compare with the case of Co adsorption, Fig.\ \ref{CoMnMgn-fig}, second row. The structure of HX-A has a reflection
 symmetry through the top-left to bottom-right diagonal as represented in Fig.\ \ref{CoMnMgn-fig}. The T-U CDW structure is modified slightly in the
 vicinity of the adsorption site. The magnetisation density in all of the CDWs is negative around Mn, unlike the case of Co, where a negative cloud
 surrounding Co is neighboured by a positive cloud around the closest Nb atoms, compare also Figs.\ \ref{CovsMnMgn-fig} (a) and (b) in the Appendix,
 showing a detailed view on the vicinity of the adsorption sites.

  Charge density differences in direct space are analysed in comparison with Co adsorption. The main difference between the two adsorbates is that with Mn adsorption
 the out-of-plane charge modulations are not suppressed. In fact, note that the in-plane red lobes in Fig.\ \ref{CoMnChgDiffs-fig} (e) replace the out-of-plane red
 lobes in Fig.\ \ref{CoMnChgDiffs-fig} (c), point in three directions, symmetrically. In fact, the similarity between Figs.\ \ref{CoMnChgDiffs-fig} (d) and (f)
 confirms that the HX-S has the same features in the case of Co and Mn, with blue lobes pointing out-of-plane.

\begin{figure*}[t]
\centering
 {\includegraphics[trim = 0 0 0 0,width=\textwidth,clip]{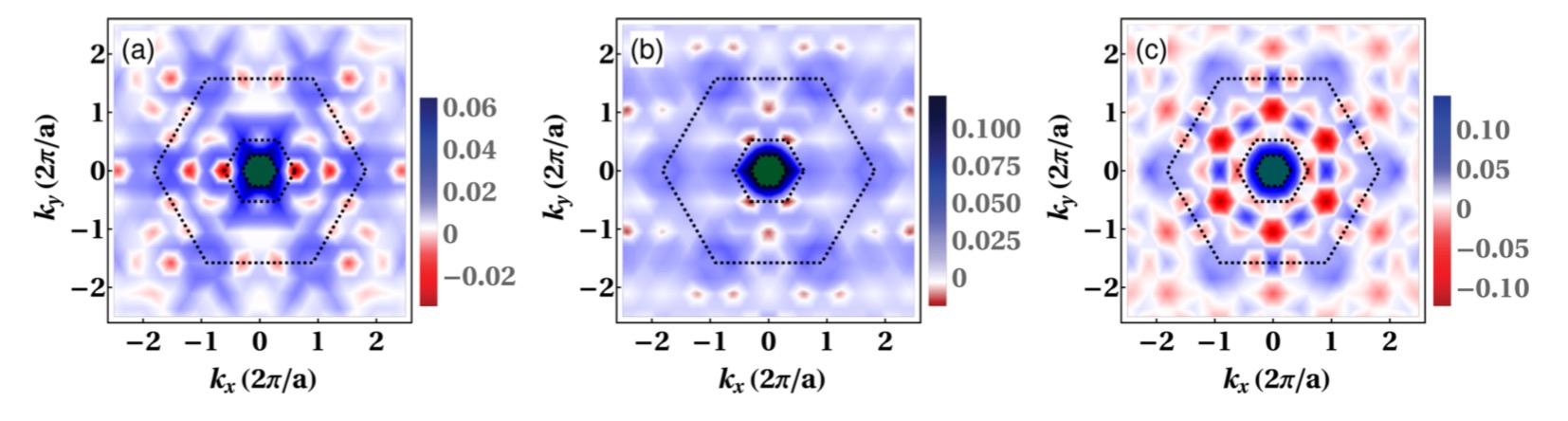}}
\caption{(Colour online) Differences of the FT of the charge density
 distributions, done with respect to pristine T-U, of Co-adsorbed (a),
 Mn-adsorbed (b) T-U, and Mn-adsorbed (c) HX-S.}
\label{otherFFT-fig}
\end{figure*}

  The analysis of the FT plots was mentioned previously with reference to Fig.\ \ref{FFTCoMn-fig}. The main difference with the case of Co adsorption is that the HX-A and HX-S
 do not differ much, i.e. the intensity variation shows little contrast between the $C_6$ and the $C_2$ symmetries, compare the HX-A and HX-S CDWs in Figs.\ \ref{FFTCoMn-fig}
 (c) and \ref{otherFFT-fig} (c). The preference of the TM for the HX CDWs seems to be at variance with their tendency of reducing the symmetry from $C_6$ to $C_2$. The lowest
 energy solutions for Co$\vert$NbSe$_2$ (and Mn$\vert$NbSe$_2$) are mixed state of HX and T-C CDWs, compare Fig.\ \ref{CoStrcts-fig} (b) with Fig.\ \ref{prstStrcts-fig} (b).
 However the Co$\vert$NbSe$_2$ HX-A CDW solution has a consistent T-C component, whereas the Mn$\vert$NbSe$_2$ HX-A CDW solution has a small T-C component, see the underlying
 structure in the magnetisation density plot, Fig.\ \ref{CoMnMgn-fig} (e).

  The CDW solutions for TM adsorbates on NbSe$_2$ are in fact mixed states; where the mixing between HX and T-C is high, the symmetry of the charge density
 distribution is reduced, whereas predominant HX CDW solutions keep a $C_6$ symmetry. In the case of Mn, the preference for HX solutions is higher than in
 the case of Co, and as a result the T-C component in the mixed solutions is smaller and in turn the $C_6$ symmetry is virtually maintained. Finally, the
 observation that the T-U / T-C have a reduced symmetry, raises the question on pristine T-U/C and the symmetry of its charge density distribution, which
 has a $C_6$ symmetry; perhaps the adsorbates induces some symmetry breaking which allows the T-U to undergo a transition.

  Table \ref{tab-occTM} shows the occupancies of Co and Mn $3d$ orbitals. The environment around the TM in the NbSe$_2$ layer is trigonal
 prismatic, and therefore orbitals split into three groups due to the crystal field; these are classified according to their symmetry into
 $\mathit{e''}$, $\mathit{e'}$ and $\mathit{a'_{1}}$. The orbital splitting is valid also for Co and Mn $3d$ orbitals, but the intensity
 of the splitting is reduced because the environment is incomplete. Therefore, the spin splitting, which depends on the $l$-character of
 the orbitals is larger than or comparable to the crystal field splitting, and orbitals are ordered by increasing energy as follows: in the
 majority spin channel, $\mathit{e''}$, $\mathit{e'}$ and $\mathit{a'_{1}}$; in the minority spin channel the order of $\mathit{e'}$ and
 $\mathit{a'_{1}}$ is inverted. Furthermore, the orbital projections onto cubic (real spherical) harmonics ($d_{xy}$, $d_{yz}$, $d_{z^2 - 3r^2}$,
 $d_{zx}$ and $d_{x^2 - y^2}$, ordered by increasing $m$ value) are spin dependent. The orbitals of the eigenbasis can be written as
 $\mathit{e''}^1_{\sigma} = v_{\sigma} d_{xy} + u_{\sigma} d_{yz}$, $\mathit{e''}^2_{\sigma} = u_{\sigma} d_{zx} + v_{\sigma} d_{x^2 - y^2}$,
 $\mathit{e'}^1_{\sigma} = u_{\sigma} d_{xy} + v_{\sigma} d_{yz}$ and $\mathit{e'}^2_{\sigma} = v_{\sigma} d_{zx} + u_{\sigma} d_{x^2 - y^2}$.
 (Freedom in the choice of the basis set allows one to set $\mathit{a'_{1}} = d_{z^2 - 3r^2}$.) The $l$-character of the $\mathit{e''}^{\uparrow}$
 orbitals is close to 1 (i.e. the $u$ values are larger than the $v$ values); conversely, the $\mathit{e'}^{\uparrow}$ orbitals have a prevalent
 $l = 2$ character ($v$ values are larger); overall, this fact holds for the $\mathit{e''}^{\downarrow}$ and $\mathit{e'}^{\downarrow}$ orbitals.

\begin{figure*}[t]
\centering
 {\includegraphics[trim = 0 0 0 0,width=\textwidth,clip]{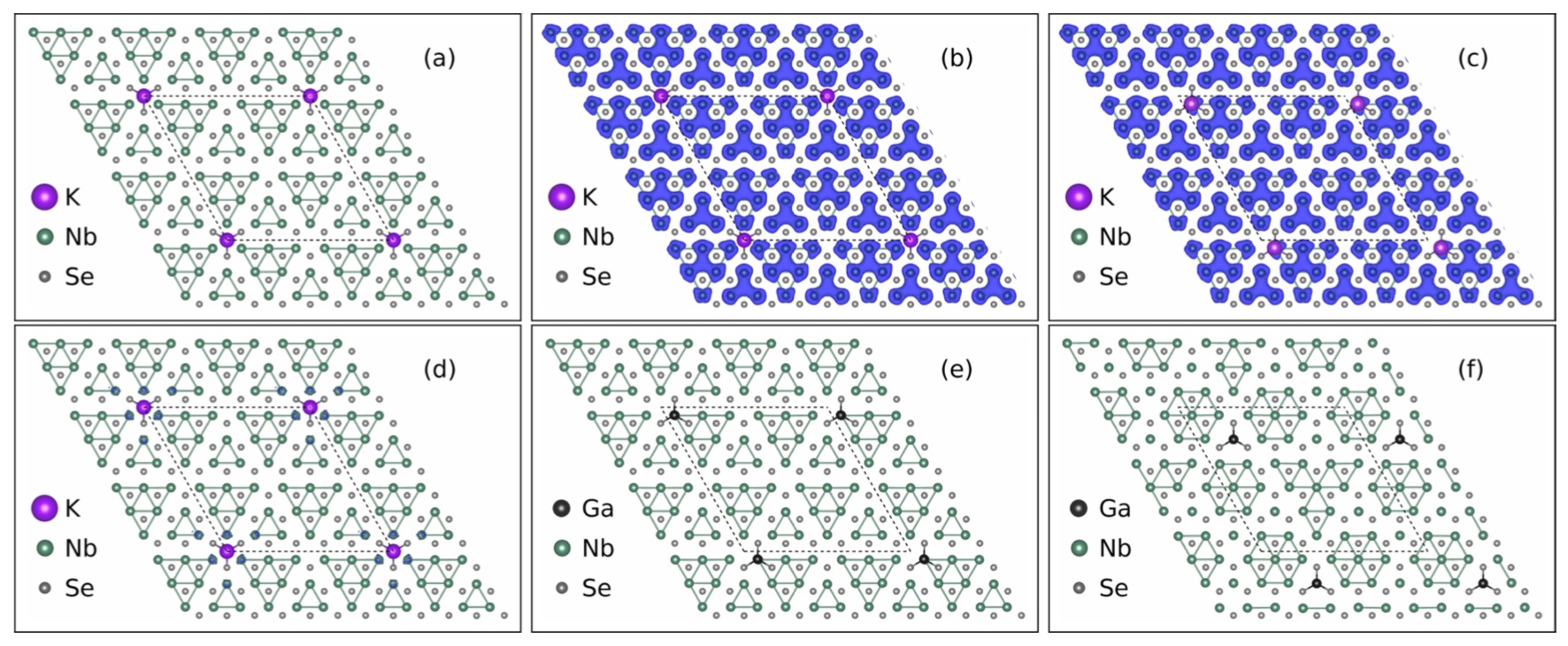}}
\caption{(Colour online) First row: ground state CDW structure, with K adsorbed on the hollow
 position (a), its charge density distribution (b), and the charge density for K adsorbed on
 the Nb site (c); Second row: Difference in the charge densities between the K$\vert$NbSe$_2$
 ground state and NbSe$_2$ T-U (d), ground state CDW structures for two Ga$\vert$NbSe$_2$,
 namely T-U (e) and HX-A (f). Atoms are represented by spheres, as illustrated in the legends;
 Nb-Nb bonds shorter than the equilibrium distance (3.45 \AA) are represented by solid lines to
 help visualising the CDW structure pattern. Dashed lines mark the supercells borders. The
 isosurface value for the volumetric data is set in agreement with Fig.\ \ref{prstStrcts-fig}.}
\label{KStrctsChg-fig}
\end{figure*}
\begin{figure*}[t]
\centering
 {\includegraphics[trim = 0 0 0 0,width=\textwidth,clip]{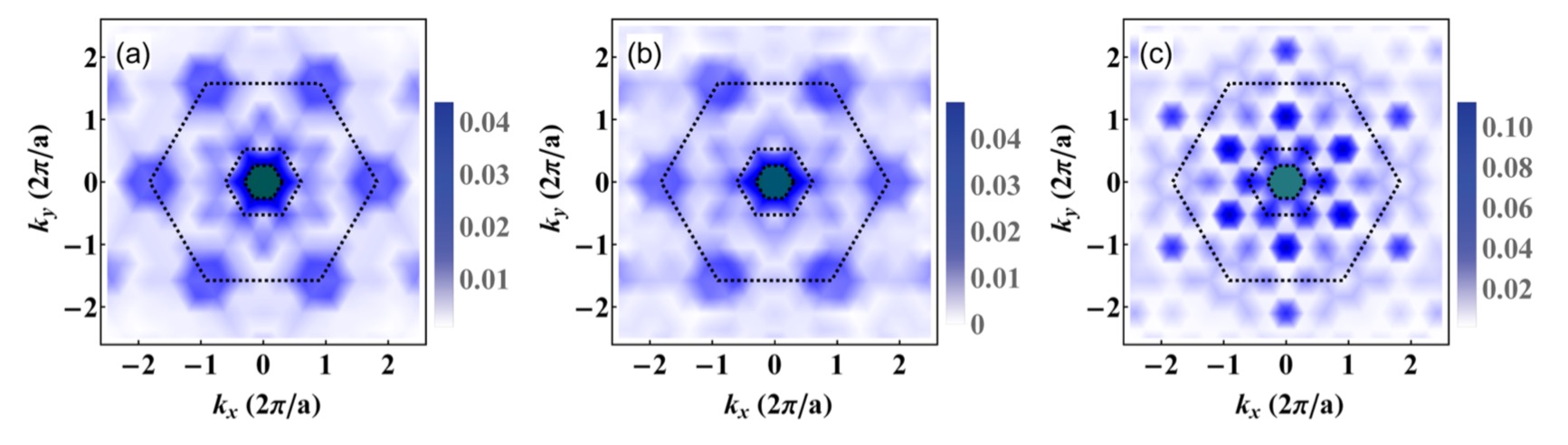}}
\caption{(Colour online) Difference in the FT of the charge density
 distributions between the K$\vert$NbSe$_2$ on the hollow site (a),
 and Nb site (b) and the T-U pristine CDW. The charge distribution has
 a $C_2$ symmetry for the adsorption on the hollow site and a $C_6$
 symmetry for the adsorption on the Nb site. In (c), the FT difference
 between K$\vert$NbSe$_2$ on the hollow site and the pristine T-C CDW
 is shown.}
\label{FFTK-fig}
\end{figure*}

  In summary, TM adsorption favours HX CDW, weakening the CDW signals for the T-U and T-C CDWs, and the symmetry of the charge distribution is reduced from
 $C_6$ to $C_2$, especially in the case of Co, where mixing between HX and T-C occur. The modulation of the magnetic density is dependent on the symmetry of
 the charge distribution: the higher the symmetry, the weaker the modulation. Also, due to a weaker crystal field splitting, the orbital order of the $3d$
 of Co and Mn in the two spin channels is different. Finally, a probe for Mn and Co is given in terms of $l$-character of their electrons.

\subsection{Adsorption of K and Ga}

 The adsorption of K is different from the other cases, since the energy difference between having an adatom 
 at the hollow site or in top of a Nb site (see Table \ref{tab-ads}) is small, and therefore can be reversed
 by the presence of a CDW. This is, in fact, what happens. While on a non modulated NbSe$_2$ structure, the
 hollow site stands 4 meV above the Nb adsorption site, on the CDW modulated structures the hollow site becomes
 more favourable, of about 0.2 meV/f.u. (their respective ground state are compared). The solutions obtained
 starting from the T-C converge to mixed states between T-C itself and HX and they are found at a high energy,
 while the T-U CDW is still favoured by 0.7 meV/f.u. over the HX CDW (the energy difference slightly changes
 with respect to the pristine case). The CDW structure and charge density distribution (in the T-U) on the
 hollow site (the ground state) and on the Nb site look very similar, see Fig.\ \ref{KStrctsChg-fig}, and trace
 back to the pristine CDW ground state, compare Fig.\ \ref{prstStrcts-fig}. The CDW is slightly enhanced inside
 the hexagon delimiting the CDW peaks, but without a leading $\mathbf q$. The symmetry of the FT plot for K on
 the Nb (hollow) site is reduced from C$_6$ to C$_2$ at the points
 $\mathbf q = (\pm \pi/3a,\pi/\sqrt{3}a)$ ($\mathbf q = (\pm 2 \pi/3a,0)$), see Fig.\ \ref{FFTK-fig}; however,
 the intensity is one order of magnitude smaller than that of the Nb and Mn cases. Note also the difference with
 the T-C CDW, showing that the signal of K-adsorbed T-U on the hollow site is significantly enhanced with respect
 to pristine T-C at distinct $\mathbf q$ vectors, see Fig.\ \ref{FFTK-fig} (c).

  The case of Ga is interesting in comparison with K because Ga has a fully occupied $s$ shell and a single electron in the $p$ shell. In this case, the adsorption on the
 Nb site is favoured by 183 meV over the adsorption at the hollow site. The T-U CDW is in strong competition or coexists with the HX CDW, and the T-C converges to a mixed
 solution between T-C and HX, analogously to the case of K. The ground state structures are shown in Fig.\ \ref{KStrctsChg-fig}. The charge density distributions and their
 FTs are not particularly different from the case with K adsorption. Both K and Ga cases are related to gate doping, because they consist of an electron per 36 f.u. injected
 in the system. However, structural reconstructions due to the chemical adsorption must be considered for the enhancement or suppression of the CDW order. Compared to
 experimental observations with gate doping \cite{xi-PRL.117.106801}, where the case of K and Ga adsorption suppress the CDW order, the FT plots suggest that a considerable
 order remains (no depletion is seen at any $\mathbf q_{CDW}$). Therefore, a doping with a hole carrying atom could enhance the CDW signal.

  Overall, adsorption of atoms on single-layers NbSe$_2$ suppresses the T-C CDW and promotes the HX CDWs in all cases; in particular, with Co and Mn the HX CDWs become
 the ground state for all the coverage considered in this study, whereas in the case of Ga the HX CDWs are at the same energy of the T-U CDW. Adsorption of K does not
 change the ground state (T-U) but does suppress the T-C CDW; the HX CDW solutions are the closest ones to the ground state (0.7 meV/f.u. above it), within thermal
 fluctuations, and therefore they are very likely to be seen by Scanning Tunnelling Microscopy in real samples. In fact, as several STM data are becoming available, a
 guide on the CDW hierarchy may be very useful to correctly identify and locate metallic impurities in TMDCs. 
 
\begin{figure*}[t]
\centering
 {\includegraphics[trim = 0 0 0 0,width=\textwidth,clip]{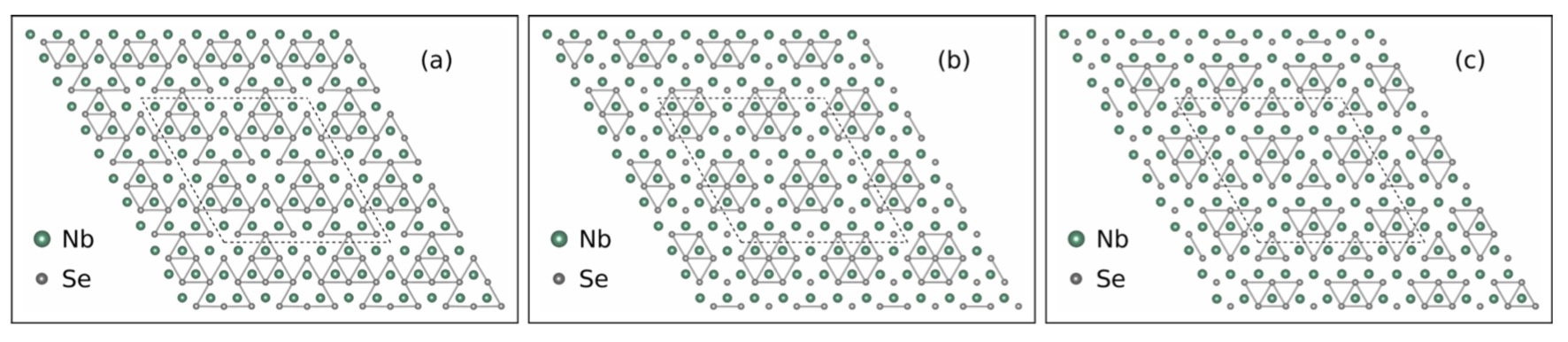}}
\caption{(Colour online) Relaxed structures of the pristine CDWs as they appear in Fig.\
 \ref{prstStrcts-fig}:
 1\ in the main text: T-U (a), T-C (b) and HX (c). Atoms are represented by
 spheres, as illustrated in the legends; Se-Se bonds shorter than the equilibrium distance
 (3.45 \AA) are represented by solid lines, in order to help visualising the CDW structure
 pattern. Dashed lines mark the supercells borders.}
\label{prstStrctsipSe-fig}
\end{figure*}
\begin{figure}[t]
\centering
 {\includegraphics[trim = 0 0 0 0,width=\linewidth,clip]{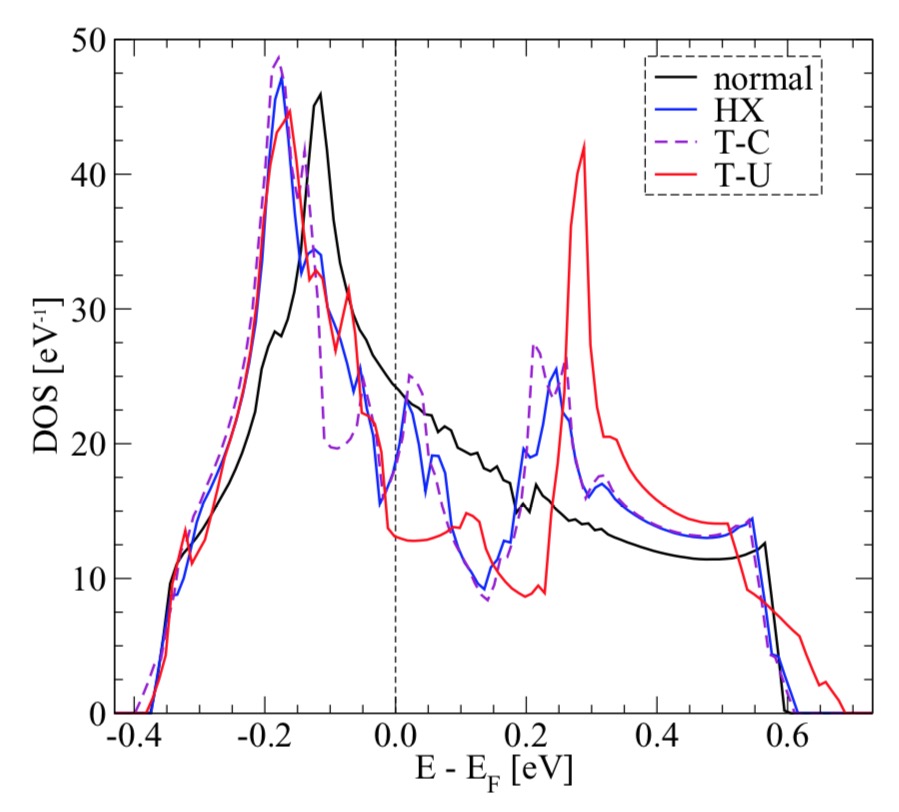}}
\caption{(Colour online) Total DOS of NbSe$_2$ with the fully symmetric
 structure and with the HX, the T-C and the T-U CDW structures, obtained
 for a $3\times3\times1$ supercell.}
\label{CDWDOS-fig}
\end{figure}
\begin{figure*}[t]
\centering
 {\includegraphics[trim = 0 0 0 0,width=\textwidth,clip]{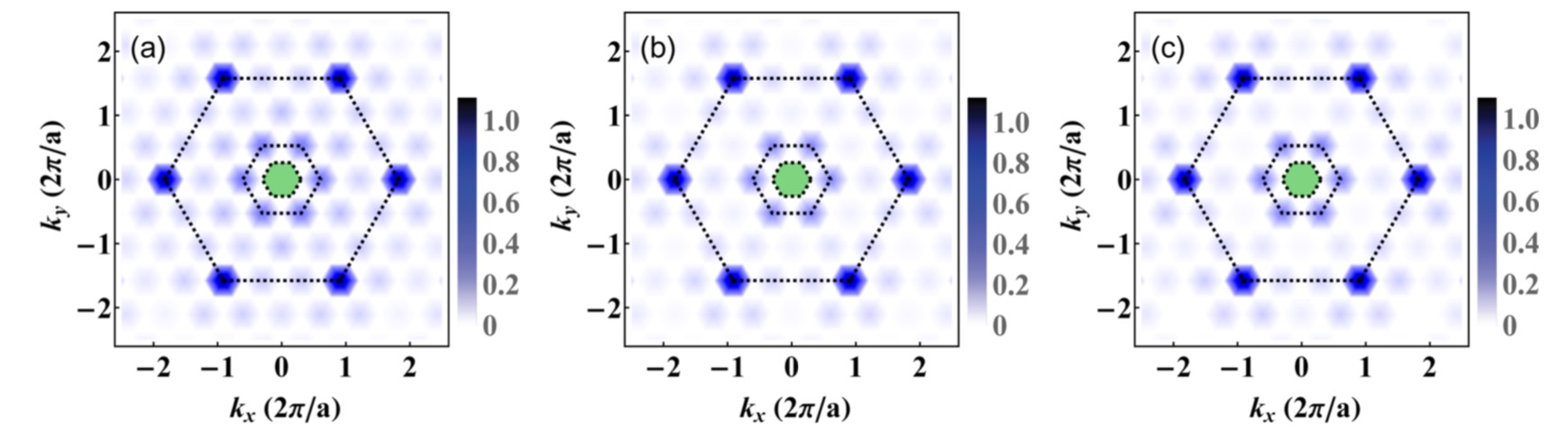}}
\caption{(Colour online) Fourier Transform (FT) plots of the charge density distributions for the
 pristine T-U (a) T-C (b) and HX (c) CDW structures, as given in Fig.\ \ref{prstStrcts-fig}. The
 dashed hexagons mark the same regions as in Fig.\ \ref{FFTprst-fig} (see caption).}
\label{prstFFT-fig}
\end{figure*}
\begin{figure*}[t]
\centering
 {\includegraphics[trim = 0 0 0 0,width=\textwidth,clip]{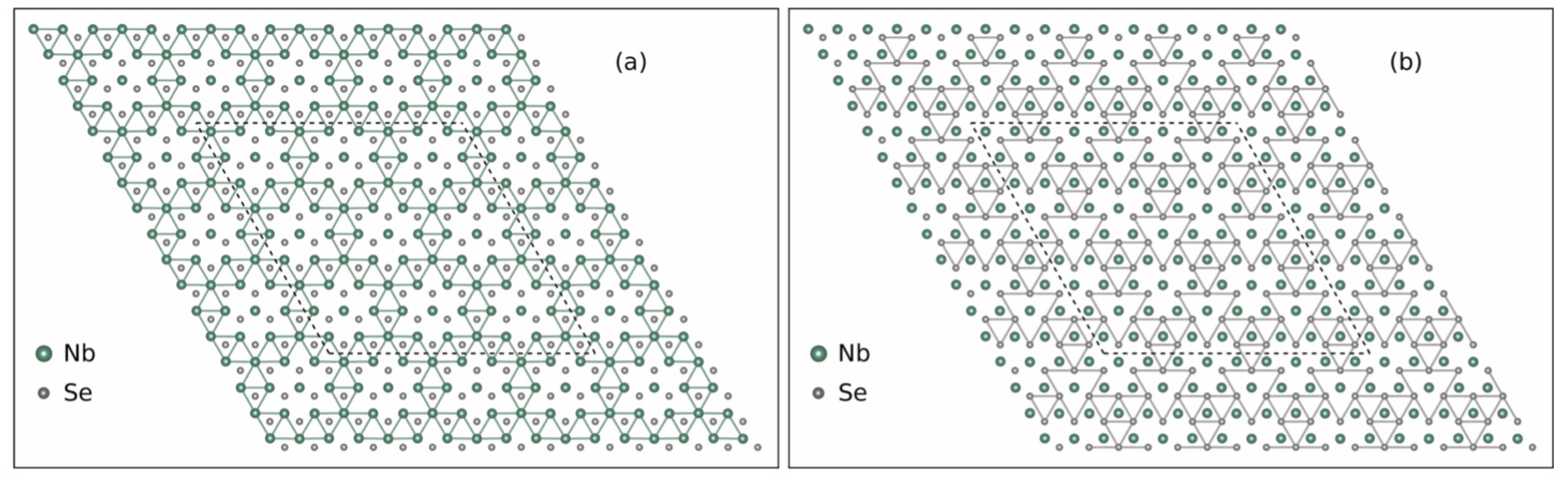}}
\caption{(Colour online) Relaxed structures of a variant of the HX CDW in a $9\times9\times1$
 supercell, highlighting its Nb-Nb (a) and Se-Se (b) distance pattern. Atoms are represented
 by spheres, as illustrated in the legends; in (a) and (b) respectively, Nb-Nb or Se-Se bonds
 shorter than the equilibrium distance (3.45 \AA) are represented by solid lines, in order to
 help visualising the CDW structure pattern. Dashed lines mark the supercells borders.}
\label{HXoreStrcts9x9-fig}
\end{figure*}
\begin{figure*}[t]
\centering
 {\includegraphics[trim = 0 0 0 0,width=\textwidth,clip]{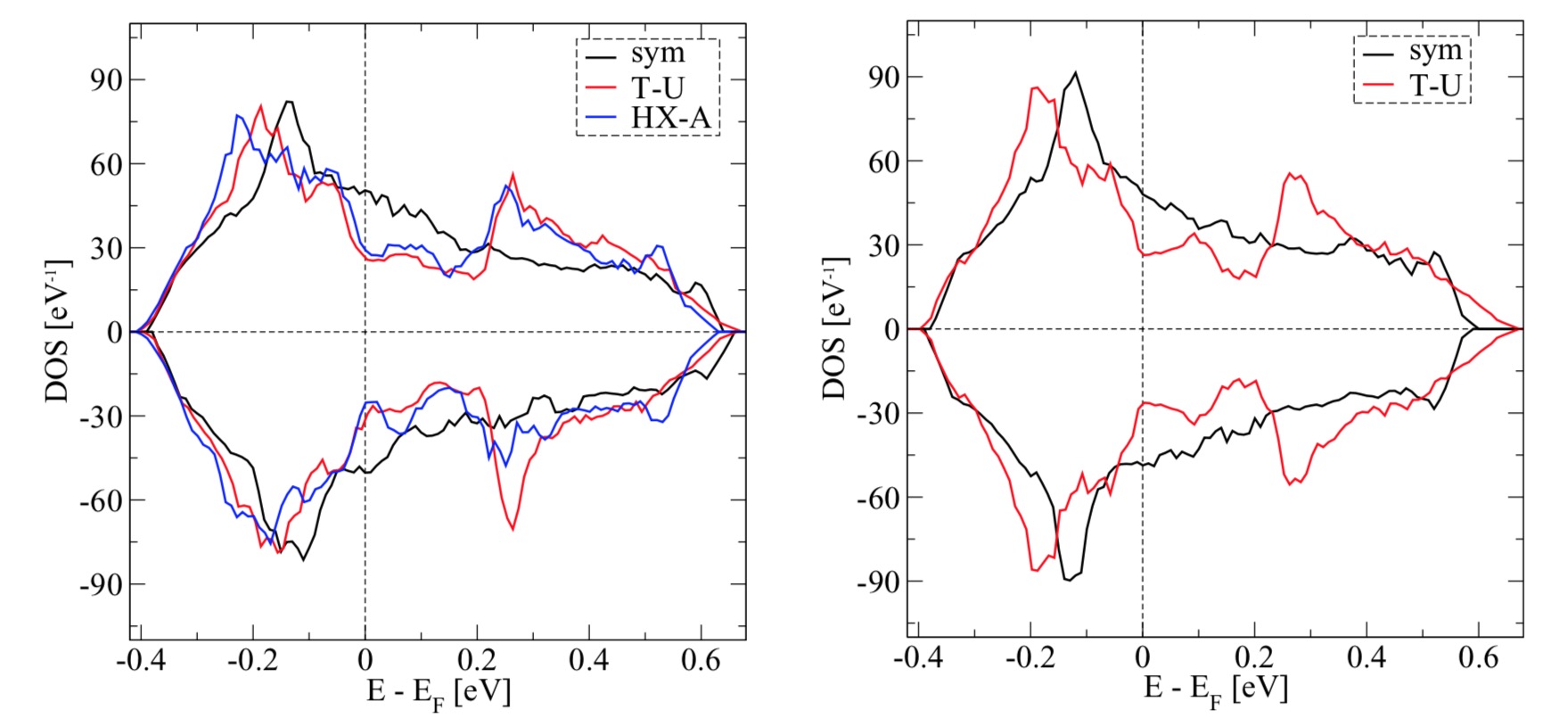}}
\caption{(Colour online) Total DOS of Co$\vert$NbSe$_2$ and K$\vert$NbSe$_2$
 with symmetric structures and with their CDW structures.}
\label{TMCDWDOS-fig}
\end{figure*}
\begin{figure*}[t]
\centering
 {\includegraphics[trim = 0 0 0 0,width=\textwidth,clip]{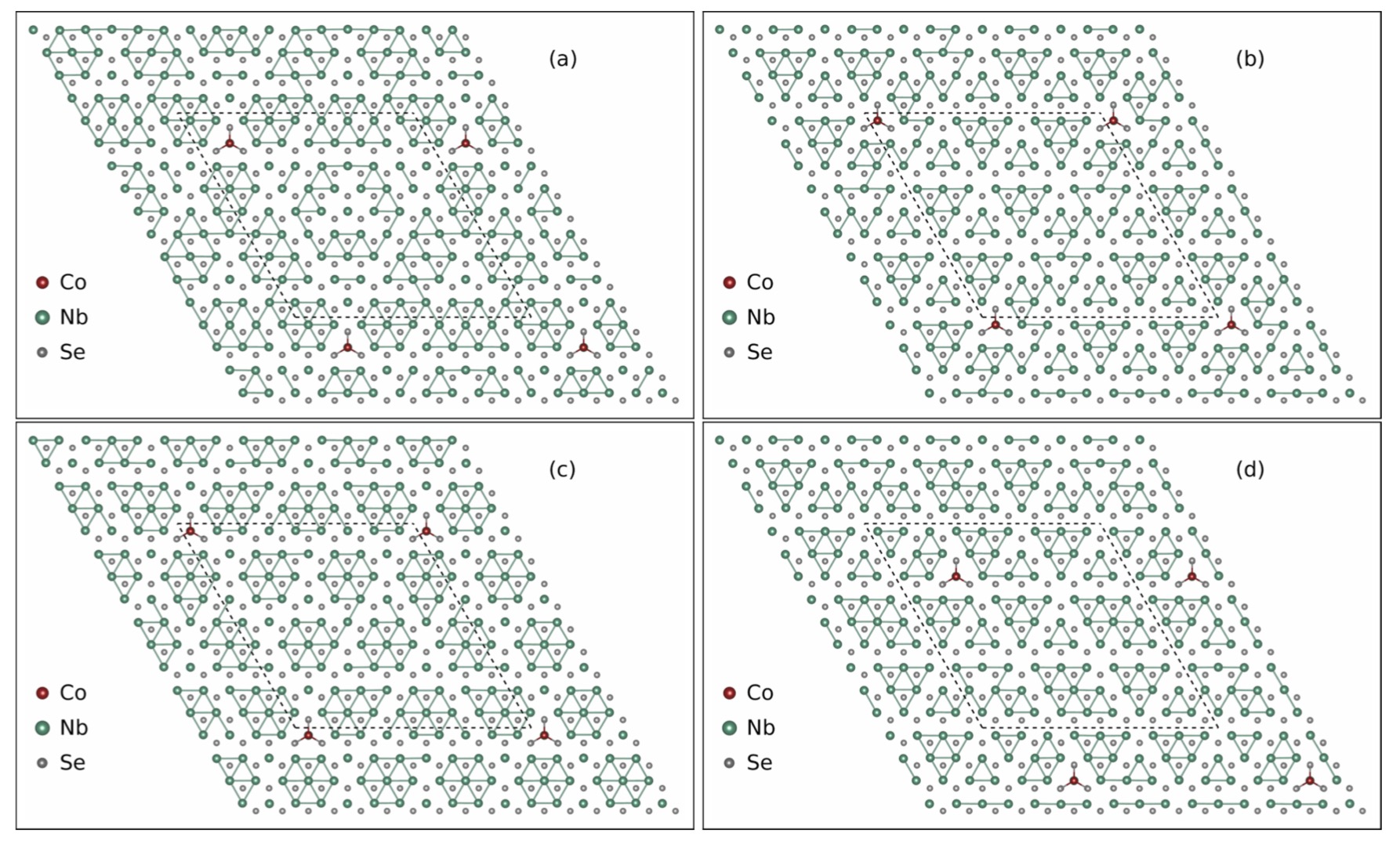}}
\caption{(Colour online) Relaxed structures of Co$\vert$NbSe$_2$ in a $9\times9\times1$
 supercell. Atoms are represented by spheres, as illustrated in the legends; Nb-Nb bonds
 shorter than the equilibrium distance (3.45 \AA) are represented by solid lines, in
 order to help visualising the CDW structure pattern. Dashed lines mark the supercells
 borders.}
\label{CoStrcts9x9-fig}
\end{figure*}
\begin{figure*}[t]
\centering
 {\includegraphics[trim = 0 0 0 0,width=\textwidth,clip]{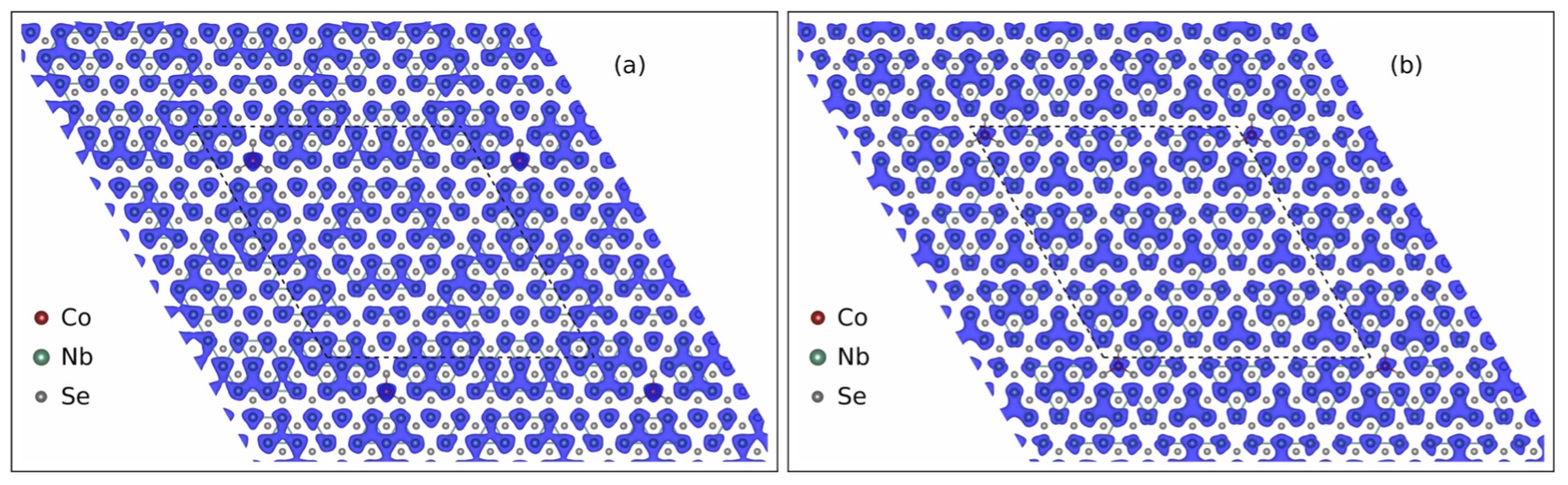}}
\caption{(Colour online) Direct space charge density distribution of Co$\vert$NbSe$_2$
 in a $9\times9\times1$ supercell; (a) and (b) correspond to the respective structures
 in Fig.\ \ref{CoStrcts9x9-fig}. The isosurface value for the volumetric data is set in
 agreement with the relative figures in the main text. Atoms are represented by spheres,
 as illustrated in the legends; Nb-Nb bonds shorter than the equilibrium distance (3.45
 \AA) are represented by solid lines, in order to help visualising the CDW structure
 pattern. Dashed lines mark the supercells borders.}
\label{CoChg9x9-fig}
\end{figure*}
\begin{figure*}
\centering
 {\includegraphics[trim = 0 0 0 0,width=\textwidth,clip]{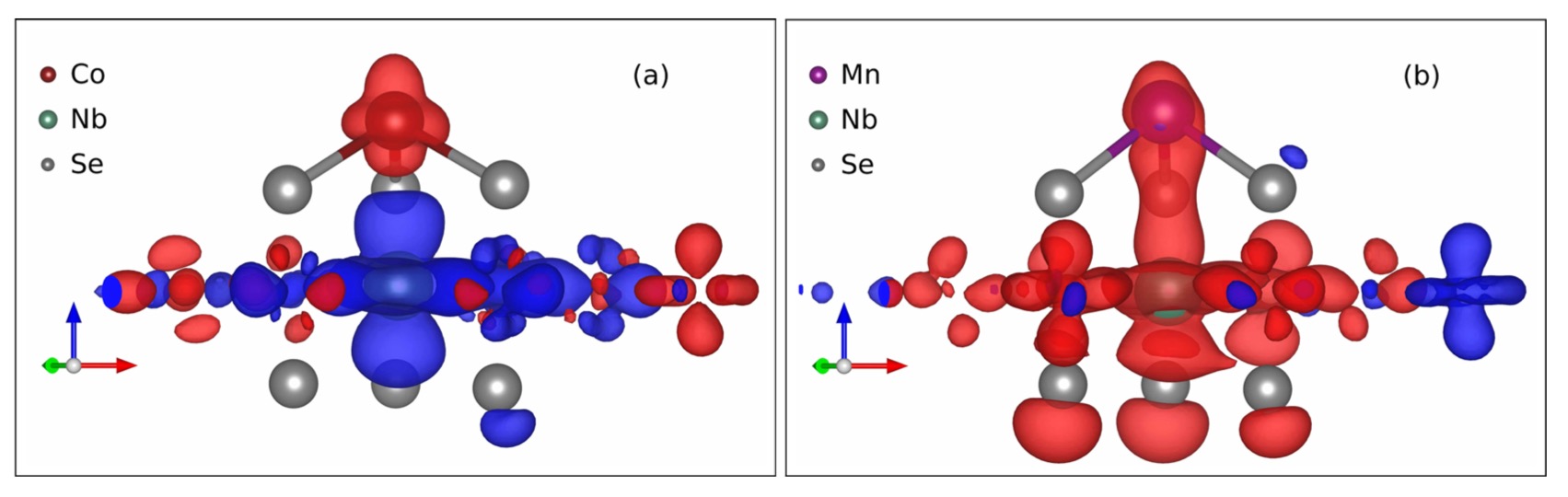}}
\caption{(Colour online) Magnetisation densities of Co$\vert$NbSe$_2$ T-U (a) and
 Mn$\vert$NbSe$_2$ T-U (b) with details around Co and Mn, respectively. The sets of
 axes represent the unitary lattice vectors in red, green and blue colour, respectively.
 Volumetric data representation are set in agreement to the relative figures in the main
 text.}
\label{CovsMnMgn-fig}
\end{figure*}

 A detailed comparison between theory and experiment requires also an analysis of the role of the substrate,
 which may induce a variety of effects, as e.g. in-plane strain and charge transfer. Recently, single layers
 were grown on bi-layer graphene, and a CDW order slightly weaker compared to the bulk was reported
 \cite{ugeda-nphys2016}; however, the phase competition between different modulations was not investigated.
 Our study suggests, based on the few examples analysed, that one CDW phase (T-U) is suppressed with respect
 to other hidden ones (HX). As a concluding remark, we observe that the symmetry of the charge density
 distribution is reduced from $C_6$ to $C_2$, hinting to a weakening of a $\mathbf q$-vector, which may be the
 precursor of a stripe phase recently observed by STM \cite{fang-ScienceAdv2018,soumyanarayanan-PNAS2013}.

\section{Conclusions}
\label{conc}
  By means of {\it ab-initio} calculations based on total energy and direct space charge computation,
 we have investigated the existence and competition of CDWs in single-layer NbSe$_2$ without and with
 impurities. The T-C CDW is suppressed in all cases, suggesting that its observation in STM images is
 unlikely in non-passivated samples, due to the high reactivity of NbSe$_2$ and, in general, metallic
 TMDCs. Transition metal adsorbates invert the energy hierarchy between CDWs, favouring the HX CDWs
 over the T-U and the T-C. Adsorption of K keep the T-U CDW as the ground state, although the HX CDW
 is preferred to the T-C CDW; adsorption of Ga equally favours the T-U and the HX CDWs pointing to a
 coexistence. In general, adsorption of atoms, changing the local symmetry mixes the `pristine' CDWs,
 in particular the HX and the T-C. The symmetry of the charge density distribution is reduced from
 $C_6$ to $C_2$ upon Co or Mn adsorption. Future research will be focused on understanding the role
 of the substrate in the stabilisation of the CDWs, in order to have a better correspondence 
 between theory and experiment.

\section*{Acknowledgements}
 We are grateful to B.\ I.\ Min, V.\ Fiorentini, E.\ Tosatti and D.\ Payne for fruitful
 discussions.
  This research work was supported by the Ministry of Education, Gyeongsangbuk-do and Pohang City, through the National Research Foundation of Korea (Grant Nos.:
 2015R1C1A1A01052411 and 2017R1D1A1B03033465).
 K.\ K. and A.\ A. acknowledge the Max Planck POSTECH / KOREA Research Initiative 
 programs through  the National Foundation of Korea (NRF) funded by the Ministry of Science, ICT and Future Planning (Grant No. 2016K1A4A4A01922028).
  The computational work was performed on resources provided
 by the Swedish National Infrastructure for Computing (SNIC) at the High Performance Computing Center
 North (HPC2N), at the PDC center for High Perfomance Computing and on the National Computational
 Infrastructure (NCI) of Australia. Further computational resources were provided by the IBS centre at
 POSTECH and the Korean Institute of Science and Technology Information (KISTI).

\appendix

\section{Pristine}
  As reported in the literature, the (multi-dimensional) potential energy surface of the CDW phase features several (rather shallow) minima; as a result, several
 symmetries are possible. The existence of two forms of orthorhombically distorted structures was mentioned in the main text, and is extensively explained in the
 literature. Some differences identified in the main text included the mention of the accompanying Se-Se bonds. Figure \ref{prstStrctsipSe-fig} (a), (b) and (c)
 show the T-U, T-C and HX CDWs respectively, highlighting their Se-Se distance patterns. Also hexagonally distorted structures can exist in two forms, one having
 hexagonal Nb-Nb patterns, analysed throughout the manuscript, and one having three-fold symmetric Nb clusters, as shown in Fig.\ \ref{HXoreStrcts9x9-fig}. This
 latter structure does not converge within our $6\times6\times1$ supercells, and appears to be stabilised only in a $9\times9\times1$ supercell; it is found at a
 higher energy with respect to the other three CDW structures (2.7 meV above the T-U CDW structure) and therefore further analysis has been dismissed; Figs.\
 \ref{HXoreStrcts9x9-fig} (a) and (b) show its structure with its Nb clusters and its Se clusters, respectively.

  With reference to Fig.\ \ref{CDWDOS-fig}, the DOS of the main three CDW structures are compared to the DOS of the fully symmetric structure of pristine NbSe$_2$. As
 presented elsewhere \cite{calandra-PRB.80.241108}, for single-layers a single band crosses the Fermi level and extends from $\sim -0.4$ eV to $\sim +0.7$ eV. As the
 main contribution for this band come from Nb states, we refer to such band as Nb band. The CDW formation shifts the spectral weight from $\sim-0.10$ eV to $\sim-0.20$
 eV and enhances it in the range ($+0.25,+0.30$) eV; the DOS peaks are aligned around $-0.20$ eV, whereas the DOS peaks around $+0.25$ eV are at different energy for
 different CDWs. The T-U CDW DOS features a depletion of spectral weight in correspondence of the Fermi level till $\sim+0.10$ eV and a trough at $\sim+0.20$ eV (which
 in fact precedes the peak at $\sim+0.30$ eV); the T-C CDW DOS has a trough at $\sim-0.10$ eV, a second one at Fermi level and a third one at $\sim+0.15$ eV; the HX
 CDW DOS has one trough at Fermi level and one at $\sim+0.15$ eV, being more similar to the T-C CDW DOS. The T-U differs from the other two also by a slightly larger
 band-width, whose tail reaches $\sim$ 0.7 eV. Overall, the CDW affects different energy ranges of the DOS both around the Fermi level and away from it, in agreement
 with the literature \cite{shen-PRL.99.216404,calandra-PRB.80.241108}.

\section{Co adsorption}

  The DOS of symmetric and selected CDW structures obtained in the $6\times6\times1$ supercells, computed and analysed for Co$\vert$NbSe$_2$ and for
 K$\vert$NbSe$_2$ in $6\times6\times1$ supercells, is shown Fig.\ \ref{TMCDWDOS-fig}. By relaxing the M$\vert$NbSe$_2$ (M = Co, K) structures fixing
 the in-plane coordinates of the Nb and Se atoms, the symmetry of each structure was preserved; their relative DOS curves are labelled `sym'. The main
 features of the DOS of the T-U and HX CDWs are maintained with respect to the pristine CDWs: the peaks at $\sim -0.20$ eV and $\sim +0.25$ eV, and the
 depletion of states around Fermi level. However, the position of the peaks for the HX-A and T-U CDWs are now different in the case of Co adsorption;
 also, while the depletion of states in the pristine HX CDW is accompanied by peaks and troughs near Fermi level, the depletion of states in the HX-A
 CDW (i.e. with Co adsorbates) shows a profile in line with the T-U CDW. In summary, the effect of the CDW on the DOS slightly differs passing from the
 pristine case to the case with adsorbates, the states around Fermi level being the most affected ones.

  With reference to Fig.\ \ref{otherFFT-fig}, we analyse further details of the effect of Co/Mn adsorption on the CDW intensity, started with Fig.\ \ref{FFTCoMn-fig}.
 The asymmetry of the FT plots, highlighted by computing the difference with the charge distribution in the pristine CDWs (T-U in the case of Fig.\ \ref{FFTCoMn-fig})
 is clearly due to the adsorbate, rather than the asymmetric nature of the CDW itself. In fact, Figs.\ \ref{otherFFT-fig} (a) and (b) show the FT of the charge densities
 of Co-adsorbed T-U and Mn-adsorbed T-U minus pristine T-U - while Fig.\ \ref{otherFFT-fig} (c) shows the Mn HX-S minus pristine T-U for comparison with the figure in
 the main text. The effect of Co adsorption and Mn adsorption on the CDW peaks is somehow opposite to each others: Co adsorption suppresses the CDW at
 $\mathbf q = (\pm 2\pi/3a, 0), (\pm 4\pi/3a, 0)$, while enhancing it along the lines $\mathbf q = (\pm \pi/a, \sqrt{3}\pi/a)$; Mn adsorption suppresses the CDW at
 $\mathbf q = (\pm \pi/3a, \pi/\sqrt{3}a), (0, 2\pi/\sqrt{3}a)$, enhancing it only at small values of $\left|\mathbf q\right|$. The asymmetry of the charge distribution
 of T-U suggests that such asymmetry is still reduced even for low concentrations of Co/Mn, when the energy competition starts to turn in favour of the T-U CDW.

  A single Co adsorbed on NbSe$_2$, with a concentration of 1 atom over 81 unit cells (a $9\times9\times1$ replica), was studied in order to assess how fast the
 energy order changes to recover that of pristine system. Four solutions were found within 0.4 meV/f.u, see Fig.\ \ref{CoStrcts9x9-fig}; such small difference in
 energy makes them accessible by thermal fluctuations. The structure of the HX solutions shows some similarities to the case with higher adsorbate coverage. Among
 the low energy solutions, the T-C CDW was found with some mixing with HX and T-U, see Figs.\ \ref{CoStrcts9x9-fig} (a), (c) and (d). The T-U solutions are closer
 in energy to the ground state solution HX-A (0.3 - 0.4 meV/f.u.), whereas other mixed structures are found at higher energy, Figs.\ \ref{CoStrcts9x9-fig} (b) and
 (d), respectively.

  The direct space charge distributions of the lowest energy HX solution and the first T-U solution are shown in Figs.\ \ref{CoChg9x9-fig} (a) and (b),
 corresponding to the structures shown in Fig.\ \ref{CoStrcts9x9-fig} (a) and (b); the HX solution shows characteristic patches of the T-C and HX CDWs,
 following the structural modulation and confirming the tendency for these two CDW to mix together upon Co adsorption; in fact, patches as represented
 in Fig.\ \ref{prstStrcts-fig} are found in the $9\times9\times1$ supercell. The charge distribution of the T-U solution looks almost identical to that
 in the $6\times6\times1$ supercell, including the reduced symmetry, which is likely due to unidirectional local structural distortions driven by the Co
 electronic degree of freedom.
 
 Finally, the energetic stability of CDWs in Co$\vert$NbSe$_2$ was also used to estimate how small changes of $U$ and $J$ affect the
 results presented in this manuscript. Several DFT+U calculations were performed for the $6\times6\times1$ supercell, sampling $U$ in
 the range $3 - 5$ eV and $J$ in the range $0.5-0.9$ eV. We found that the values reported in Table \ref{tab-cdwia} are independent
 from the chosen Coulomb interaction parameters. The only exception is the calculation for $U=5$ eV and $J=0.5$ eV, when the HX-A CDW
 ends up in a different solution (minimum), much higher in energy, while the other two CDWs keep their energetic arrangement.

\section{Magnetic density}
  The comparison between magnetic couplings within the Nb band in the two cases of Co adsorption and Mn adsorption has been discussed in the main text. The actual magnetic
 moments (i.e. involving all states up to Fermi level) of Co, Mn and their respective underlying Nb atoms are $+1.9 \mu_B$, $+4.4 \mu_B$, $+0.6 \mu_B$ and $-0.8 \mu_B$,
 respectively. In summary, the Co and Mn magnetic moments in the Nb bands are inverted with respect to the total magnetic moments, while the underlying Nb magnetic moments
 do not change sign, remaining positive in the case of Co adsorption and negative in the case of Mn adsorption.


%
\bibliography{biblio_FC.bib}
%
\end{document}